\documentclass[structabstract]{aa}
\usepackage{natbib}
\usepackage[T1]{fontenc}
\usepackage[version=3]{mhchem}
\usepackage{times}

\usepackage{color}

\usepackage{amsmath}

\usepackage[normalem]{ulem}
\usepackage{cancel}

\usepackage{txfonts}

\usepackage{graphicx}

\DeclareMathAlphabet{\mathpzc}{OT1}{pzc}{m}{it}
\usepackage{epstopdf}

\newcommand{\CII}{[C\,{\sc ii}]}   
\newcommand{\CI}{[C\,{\sc i}]}   
\newcommand{\OI}{[O\,{\sc i}]}

\newcommand{\msol}{M$_{\odot}$}   

\newcommand{\ktau}{\textrm{KOSMA-}$\mathrm{\tau}$}

\newcommand{\emm}[1]{\ensuremath{#1}}   
\newcommand{\emr}[1]{\emm{\mathrm{#1}}} 

\renewcommand{\deg}{\emm{^\circ}}

\newcommand{\Av}{\emm{\emr{A_v}}} 

\newcommand{\HH}   {\mbox{H$_2$}}       
\newcommand{\Cp}   {\mbox{C$^{+}$}}   
\makeatletter
\newcounter{reaction}
\renewcommand\thereaction{C\,\arabic{reaction}}
\newcommand\reactiontag{\refstepcounter{reaction}\tag{\thereaction}}
\newcommand\reaction@[2][]{\begin{equation}\ce{#2}%
\ifx\@empty#1\@empty\else\label{#1}\fi%
\reactiontag\end{equation}}
\newcommand\reaction@nonumber[1]{\begin{equation*}\ce{#1}%
\end{equation*}}
\newcommand\reaction{\@ifstar{\reaction@nonumber}{\reaction@}}
\makeatother

\begin{document}
\def\aj{AJ}%
\def\actaa{Acta Astron.}%
\def\araa{ARA\&A}%
\def\apj{ApJ}%
\def\apjl{ApJ}%
\def\apjs{ApJS}%
\def\ao{Appl.~Opt.}%
\def\apss{Ap\&SS}%
\def\aap{A\&A}%
\def\aapr{A\&A~Rev.}%
\def\aaps{A\&AS}%
\def\azh{AZh}%
\def\baas{BAAS}%
\def\bac{Bull. astr. Inst. Czechosl.}%
\def\caa{Chinese Astron. Astrophys.}%
\def\cjaa{Chinese J. Astron. Astrophys.}%
\def\icarus{Icarus}%
\def\jcap{J. Cosmology Astropart. Phys.}%
\def\jrasc{JRASC}%
\def\mnras{MNRAS}%
\def\memras{MmRAS}%
\def\na{New A}%
\def\nar{New A Rev.}%
\def\pasa{PASA}%
\def\pra{Phys.~Rev.~A}%
\def\prb{Phys.~Rev.~B}%
\def\prc{Phys.~Rev.~C}%
\def\prd{Phys.~Rev.~D}%
\def\pre{Phys.~Rev.~E}%
\def\prl{Phys.~Rev.~Lett.}%
\def\pasp{PASP}%
\def\pasj{PASJ}%
\def\qjras{QJRAS}%
\def\rmxaa{Rev. Mexicana Astron. Astrofis.}%
\def\skytel{S\&T}%
\def\solphys{Sol.~Phys.}%
\def\sovast{Soviet~Ast.}%
\def\ssr{Space~Sci.~Rev.}%
\def\zap{ZAp}%
\def\nat{Nature}%
\def\iaucirc{IAU~Circ.}%
\def\aplett{Astrophys.~Lett.}%
\def\apspr{Astrophys.~Space~Phys.~Res.}%
\def\bain{Bull.~Astron.~Inst.~Netherlands}%
\def\fcp{Fund.~Cosmic~Phys.}%
\def\gca{Geochim.~Cosmochim.~Acta}%
\def\grl{Geophys.~Res.~Lett.}%
\def\jcp{J.~Chem.~Phys.}%
\def\jgr{J.~Geophys.~Res.}%
\def\jqsrt{J.~Quant.~Spec.~Radiat.~Transf.}%
\def\memsai{Mem.~Soc.~Astron.~Italiana}%
\def\nphysa{Nucl.~Phys.~A}%
\def\physrep{Phys.~Rep.}%
\def\physscr{Phys.~Scr}%
\def\planss{Planet.~Space~Sci.}%
\def\procspie{Proc.~SPIE}%
\let\astap=\aap
\let\apjlett=\apjl
\let\apjsupp=\apjs
\let\applopt=\ao

\title{Full SED fitting with the \ktau\  PDR code - I. Dust modelling}
\titlerunning{Full SED fitting with the \ktau\  PDR code}
\author{M. R\"ollig\inst{1}, R.Szczerba \inst{2}
,V. Ossenkopf\inst{1}, C.Gl\"uck\inst{1}}
\authorrunning{R\"ollig et al. }
\institute{I. Physikalisches Institut, Universit\"at zu K\"oln, Z\"ulpicher Str. 77, D-50937 K\"oln,
Germany
\and N. Copernicus Astronomical Center, Rabianska 8, 87-100 Toru\'n, Poland}

\offprints{M. R\"ollig,\\
 \email{roellig@ph1.uni-koeln.de}}

\abstract{} 
{We revised the treatment of interstellar dust in the \ktau\ PDR (photo-dissociation region) model code to achieve a consistent description of the dust-related physics in the code. The detailed knowledge of the dust properties is then used to compute the dust continuum emission together with the line emission of chemical species.  } 
{We coupled the \ktau\ PDR code with the MCDRT (multi component dust radiative transfer) code to to solve the frequency-dependent radiative transfer equations and the thermal balance equation in a dusty clump under the assumption of spherical symmetry, assuming thermal equilibrium in calculating the dust temperatures, neglecting non-equilibrium effects. We updated the calculation of the photoelectric heating and extended the parametrization range for the photoelectric heating toward high densities and UV fields. We revised the computation of the \HH\ formation on grain surfaces to include the Eley-Rideal effect, thus allowing for high-temperature \HH\ formation. } 
{We demonstrate how the different optical properties,  temperatures, and heating and cooling capabilities of the grains influence the physical and chemical structure of a model cloud.
 The most influential modification is the treatment of \HH\ formation on grain surfaces that allows for chemisorption. This increases the total \HH\ formation significantly and the connected \HH\ formation heating provides a profound heating contribution in the outer layers of the model clumps. The contribution of polycyclic aromatic hydrocarbons (PAH) surfaces to the photoelectric heating and \HH\ formation provides a boost to the temperature of outer cloud layers, which is clearly traced by high-$J$ CO lines.
Increasing the fraction of small grains in the dust size distribution results in hotter gas in the outer cloud layers caused by more efficient heating and cooler cloud centers, which is in turn caused by the more efficient FUV extinction.

} 
{}


\maketitle

\section{Introduction}
Photo-dissociation regions (PDRs) are  interstellar neighbors of HII regions that are already shielded from extreme ultraviolet (EUV) photons with an energy sufficient to ionize atomic hydrogen, but where the physical and chemical conditions of the atomic and molecular interstellar medium (ISM) are still governed by the remaining far-ultraviolet (FUV) radiation of nearby massive stars \citep{HT99}. Here, the transition from the atomic to the molecular ISM takes place, giving rise to a rich astrochemical environment.  The combination of a chemically rich pool of species that includes a large reservoir of free electrons and strong energetic excitation conditions produces a wealth of spectroscopic emissions, carrying information on the internal chemical and physical structure of the PDR.

Comparing the results of theoretical and numerical calculations to the observed spectral line emission of PDRs is a frequently used method to infer their local physical conditions. Numerical models to describe the physical and chemical processes in molecular clouds have been successfully used for many years. They involve simultaneously solving the problem of a) radiative transfer for a cloud of gas and dust; b) the chemical structure by balancing formation and destruction processes for all included species; c) determining the thermal structure of the cloud by balancing all important heating and cooling processes. 

Properly taking into account all relevant chemical and physical processes is extremely time-consuming in terms of computational power and  usually leads to a large number of simplifying assumptions in the treatment of the physics and the chemistry. The exact choice of these simplifications depends on the field of application, on the expertise of the modeller, and on the available physical and chemical knowledge. Many processes are still not fully understood if not largely unknown. A prominent example is the description of interstellar dust \citep[see][ for a review on interstellar dust]{draine2003}. Many important physical and chemical processes require a good knowledge of the properties of interstellar dust grains \citep{abel08}. Chemical reactions on the surface of dust grains appear to be the only efficient formation route for a number of important astrochemical species \citep{garrod2008,hall2010}. The specifics of their reaction kinematics depend to a large degree on the material properties and the structure of the dust grains. Porous, spongy grains provide a large surface for chemical reactions but might hinder the release of the newly formed species into the gas phase \citep{leger1985, roberts2007,taquet2012}. Not only the material and shape, but also the size of the grains might be important. Very small grains and very large molecules, such as polycyclic aromatic hydrocarbons (PAHs), efficiently contribute to the gas heating by means of the photoelectric (PE) effect \citep{bt94, WD01PEH}  while larger grains are dominating the scattering and absorption of FUV photons. For a review on PAHs see \citet{tielens2008}.  

The rapid development in observational techniques over the last two decades left us with a vast amount of spectroscopic data that defies reproduction with simple numerical PDR models. We are far away from being able to really understand the conditions in any observed PDR \citep{comparison07}. In this paper we describe a number of steps that have been taken to increase the modeling power of the \ktau\ PDR model code. We tried to evolve the code toward a consistent treatment of dust physics and chemistry. By calculating the optical and energetic properties of a model cloud with a given dust composition we are now able to describe the full spectral emission characteristic of a model cloud including dust continuum and line emission\footnote{Other PDR model codes are also able to perform wavelength-dependent UV radiative transfer. Compare for example \citet{lepetit2006, shaw2005}.}.
\section {Code description\label{sect:code}}
\subsection{The \ktau\ PDR model code\label{sect:ktau}}
The \ktau\ code \citep{roellig06}, a welltested and mature PDR model code \citep{comparison07}, applies spherical geometry to the problem of simultaneously solving the chemical and energy balance in an interstellar molecular cloud. \ktau\ is equipped with a modular chemical network, i.e., chemical species can easily be added or removed from the network and the network will rebuild dynamically, including isotopic chemistry of \ce{^{13}C} and \ce{^{18}O}. After obtaining the physical and chemical structure of the model cloud, a radiative transfer code is applied to calculate the resulting emissivities for the spectral line emission.
Because of the spherical geometry of the model clump, radiation can reach a point inside the clump from any direction.

\subsection{Multi-component dust radiative transfer model\label{sect:mcdrt}}

The multi-component dust radiative transfer (MCDRT) code allows one to solve the frequency-dependent radiative transfer equations and the thermal balance equation in a dusty clump under the assumption of spherical symmetry \citep{yorke1980}. A detailed description of the code's main features is given in \citet{szczerba97}. The code includes isotropic scattering, which is important at high optical depths. Further modifications to that version of the code have been made. First, we have added the possibility to solve the radiative transfer simultaneously for a number of dust sorts (NDS) $i$ that may or may not be co-spatial, and obey any dust size distribution. The grain size distribution is given by the general relation valid for grains with radius  $a_{i,-}  \leq a \leq a_{i,+}$;
\begin{equation}\label{dsdlaw}
dn_i(a)=n f_i(a) da,
\end{equation}
where: $n_i(a)$ is the number density of grains with size $\leq a$ and  $n$ is the number density of H nuclei (in both atoms and molecules: $n=n_\mathrm{H}+2n_\mathrm{H_2}$). The function $f_i(a)$ has a modular structure and can be changed easily to a different form. Second, because we  adapted the code to the conditions typical for PDRs it was necessary to introduce an option for switching off the central source. In addition, because intense sources of an interstellar radiation field are ubiquitous in PDRs, we changed the spatial distribution of impact parameters (originally logarithmically spaced beginning from the inner shell of the clump)  along which the ray equations are integrated. The end result of this code is the infrared radiation flux emitted by the clump, which can be compared to the radiation observed from real PDRs. However, as a byproduct,  the code provides the mean intensity of the radiation field, $J_{\lambda}(r)$ and the dust temperature, $T_{\rm d,i}(r,a)$, which for now is computed from the assumed thermal equilibrium for each dust species and dust grain size at each radius $r$ of the clump. These two quantities (mean radiation field and  dust temperature) are now used in the \ktau\  PDR model code to obtain a self-consistent gas-dust model.

\subsection{Dust models\label{sect:dustmodels}}

For the purpose of this paper we selected three dust models of interstellar dust from \citet{WD01} (hereafter WD01), and the MRN dust model \citep{mrn} for comparison. The MRN interstellar dust model was constructed by fitting a dust model to the Galactic mean extinction curve. It consists of two separate dust populations, one for "astronomical silicates" (sil) and one for graphite (gra). In the MRN model the grain size distribution is identical for both dust components and is given by
\begin{equation}\label{MRNlaw}
d n_{\rm i}(a)=n A_{\rm i} a^{-3.5}da,    0.005 \le a \le 0.25 {\mu}m,
\end{equation}
where {\rm i} stands for sil and gra, $A_{\rm sil}=10^{-25.10}$, and $A_{\rm gra}=10^{-25.13}$ (WD01).

Recently gathered observational constraints led WD01 to construct new models for interstellar dust. The authors proposed that interstellar dust is composed of four main components: astronomical silicates, graphite, and populations of very small grains that consist of neutral (PAH$^{\rm 0}$)  and ionized  (PAH$^+$) particles. The grain size distributions for these  dust components follow the general definition given by Eq.~(\ref{dsdlaw}) and their functional forms ($f_i(a)$ in Eq.~(\ref{dsdlaw})) are given by Eqs.\,(2--6) of WD01. The authors employ a power-law form for the large grains and a log-normal distribution for the very small grain population.
 Free parameters of the $f_i(a)$ functions were determined by the best fit to the average extinction with $R_V=A_{\rm V}/E(B-V)$) of 3.1, 4 and 5.5 (see WD01 for
details).  $A_{\rm V}$ is the absolute visual extinction at 
V\,=\,5500\,{\AA} and $E(B-V)=A_{\rm B}-A_{\rm V}$.
   The three dust models from WD01 correspond to  lines 7, 21, and 25 from Table 1 in \citet{WD01} and are denoted WD01-7 ($R_{\rm V}=3.1$), WD01-21 ($R_{\rm V}=4.0$), and WD01-25 ($R_{\rm V}=5.5$).

To compute the extinction efficiency $Q_{\rm ext}$ and albedo $\omega$ for each dust component, we  assumed that grains are spherical and used the Mie theory \citep{bohren83}. For silicates and graphite we used the dielectric constants from \citet{draine2003a}, while for very small grains we followed the approach given by \citet{li01b}. The minimal and maximal grain size for the size distribution of each dust component are given in Tab.~\ref{table:1}. For very small grains we used 17 grain sizes, while for bigger grains we  divided each dex of grain sizes into ten sizes, keeping equal distance in $\log(a)$. The wavelength coverage used in the MCDRT code extends from 10 \AA\ to 3000 $\mu$m and is split into 333 wavelengths.

\section{Impact on the PDR model\label{sect:impactmodel}}

In the following we describe how the new, full radiative transfer (RT) computation compares to the old approximation used in the \ktau\ model. For this purpose we evaluated a reference model clump with varying FUV radiative transfer and different dust models. We kept the following model parameters constant for all models: the total clump mass $M=$10 \msol, the total surface gas density $n=10^5$~cm$^{-3}$, and the total FUV field $\chi=1000$ in units of the Draine field \citep{draine78}. We applied a power law density gradient with power index 1.5 with a constant central gas density for radii smaller than $R_\mathrm{tot}/5$. This implies $R_\mathrm{tot}=2.46\times 10^{17}\,\mathrm{cm}=0.08\mathrm{pc}$. We assumed a  total cosmic ray ionization rate of molecular hydrogen $\zeta_\ce{H2}=5\times 10^{-17}$ s$^{-1}$. The chemistry is based on the UMIST~2006 database for astrochemistry UDfA\footnote{www.udfa.net} \citep{udfa06}, using 47 chemical species and 490 reactions in total.

\subsection{UV continuum radiative transfer\label{sect:uvrt}}

\subsubsection{Dust extinction\label{sect:dustextinction}}

For any given line of sight, the optical depth at FUV wavelengths $\tau_{\rm FUV}=\sigma_{\rm d,FUV} N_{\rm H}$ \citep{SD89}, with $\sigma_{\rm d,FUV}$ being the effective dust grain extinction cross section in  the FUV, and the total hydrogen column density $N_{\rm H}$. 
In the previous version of the KOSMA-$\tau$ code, the local radiation field was calculated by computing the optical depth as a function of angle and then performing an angular averaging of the resulting intensity over the full solid angle. 
A detailed description of the FUV transfer in the spherical model has been given by \citet{stoerzer1996}.  

For a constant dust-to-H ratio the {\it total} extinction cross section per H nucleus can be derived from
\begin{equation}
\sigma_\mathrm{ext}(\lambda)=\frac{\tau_{\rm ext}(\lambda)}{ N_{\rm H} }=
{{\sum\limits_{i=1}^{\rm NDS}}
\int\limits_{a_{i,-}}^{a_{i,+}} 
    {f_i(a) C_{\rm ext}^i(\lambda,a) da} },  
\end{equation}
where $C_{\rm ext}^i(\lambda,a)$ is the 
extinction cross section of a single dust particle with size $a$ at wavelength 
$\lambda$.  Often the absolute extinction in magnitudes is used
\begin{equation}
A(\lambda) = 2.5 \log(e) 
\tau_{\rm ext}(\lambda) \approx 1.086 \tau_{\rm ext}(\lambda).
\end{equation}

By introducing the average mass extinction coefficient for each dust component, i.e., the mean extinction cross section per dust mass, 
\begin{equation}
\langle K_{\rm ext}^i(\lambda)\rangle_a =
\frac{\int\limits_{a_{i,-}}^{a_{i,+}} 
    {n_i(a) C_{\rm ext}^i(\lambda) da}}
{
\int\limits_{a_{i,-}}^{a_{i,+}} 
    {n_i(a) 4/3 \pi a^3 \rho_{{\rm b},i} da}  
}
=
\frac{\int\limits_{a_{i,-}}^{a_{i,+}} 
    {n_i(a) C_{\rm ext}^i(\lambda) da}}
{ \rho_{{\rm dust},i} }  ,  
\end{equation}
where $\rho_{{\rm b},i}$ is the bulk dust grain density (i.e., the density of
material of which dust of given sort is composed) and
$\rho_{\rm dust,i}$ is volume density of the $i$-th dust sort, the
optical depth at any given point of the model clump can be written as
\begin{equation}
\tau_{\rm ext}(\lambda) = 
\int dr
{\sum\limits_{i=1}^{\rm NDS}
    {\langle{K}_{\rm ext}^i(\lambda)\rangle_a \,\, \rho_{\rm dust,i}}} \;.
\end{equation}
Relating this to the gas density provides
\begin{equation}
\tau_{\rm ext}(\lambda) =  m_{\rm H} \int dr \,
{n\, \frac{\rho_{\rm dust}}{n\,m_{\rm H}}}\,
{\sum\limits_{i=1}^{\rm NDS}}
    {\langle{K}_{\rm ext}^i(\lambda)\rangle_a} 
    \frac{\rho_{\rm dust,i}}{\rho_{\rm dust}}   ,
\end{equation}
where ${n\,m_{\rm H}}\,=\,\rho_{\rm gas}$,
and $m_{\rm H}$ is mass of a single H atom. Using
 $\Psi\,=\,\rho_{\rm dust}/({n\,m_{\rm H}})$ 
as the total dust-to-H mass ratio,
and $\Psi\,=\,\sum\limits_{i=1}^{\rm NDS}\Psi_{\rm i}\,=\,
{\sum\limits_{i=1}^{\rm NDS}}\rho_{\rm dust,i}/({n m_{\rm H}})$ we can write
\begin{equation}
\tau_{\rm ext}(\lambda)  = m_{\rm H} 
\int dr \,
{n\, \Psi \, {\langle{K}_{\rm ext}(\lambda)}\rangle},\label{tauV3}
\end{equation}
with
\begin{equation}
\langle{K}_{\rm ext}(\lambda)\rangle=
\frac{1}{\Psi}\,{\sum\limits_{i=1}^{\rm NDS}}
    {\langle{K}_{\rm ext}^i(\lambda)\rangle_a}\,\Psi_{\rm i} .
\end{equation}

For a constant dust-to-H mass ratio the integration over $r$ in Eq. (\ref{tauV3}) can be performed 
giving an H column density $N_{\rm H}$ using
\begin{equation}
\tau_{\rm ext}(\lambda) =
N_{\rm H} m_{\rm H} \Psi \langle{{K}_{\rm ext}(\lambda)}\rangle,
\end{equation}
i.e., $\sigma_\mathrm{ext}(\lambda)=\langle{{K}_{\rm ext}(\lambda)}\rangle
m_{\rm H} \Psi$.

\subsubsection{Influence of the dust composition\label{sect:dustcomposition}}

\begin{figure}[t]
\resizebox{\hsize}{!}{\includegraphics[angle=0]{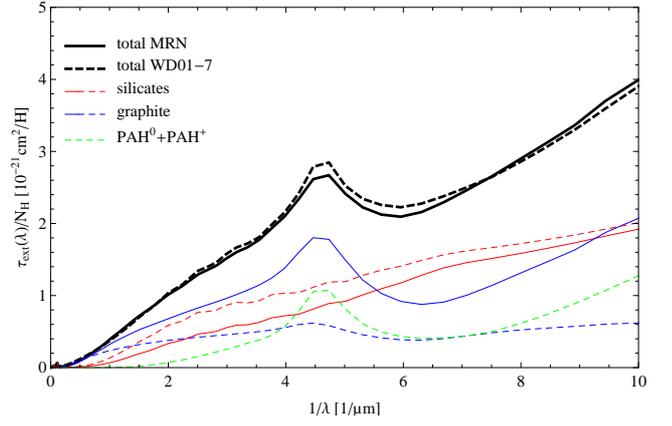}}
\caption{Extinction cross section per H nucleus  for the WD01-7 dust model (dashed lines) and for the MRN model (solid lines). The total cross sections are represented by thick lines. The contribution from each dust component is shown by red lines for silicates, blue lines for graphite, and green lines for very small grains.}\label{extcross}
\end{figure}

Figure \ref{extcross} shows the total extinction cross section per H nucleus computed for the WD01-7 (thick dashed line) and the MRN model (thick solid line) of interstellar dust. The other lines show contributions from each dust component incorporated into the WD01-7 model (silicates -- dashed, red in the electronic version; graphite -- dashed, blue in the electronic version; and PAH$^0$+PAH$^+$ --- dashed green line in the electronic version), and the solid lines present contribution from silicates (red) and graphite (blue) in the MRN model.  One can see that in the WD01 model populations of very small grains (PAH$^{\rm 0}$ and PAH$^{+}$) are responsible for a large part of the total extinction, especially for the 2200\,\AA\ band, while large grains dominate the total mean Galactic extinction in the MRN model.

\begin{figure}[t]
\resizebox{\hsize}{!}{\includegraphics[angle=0]{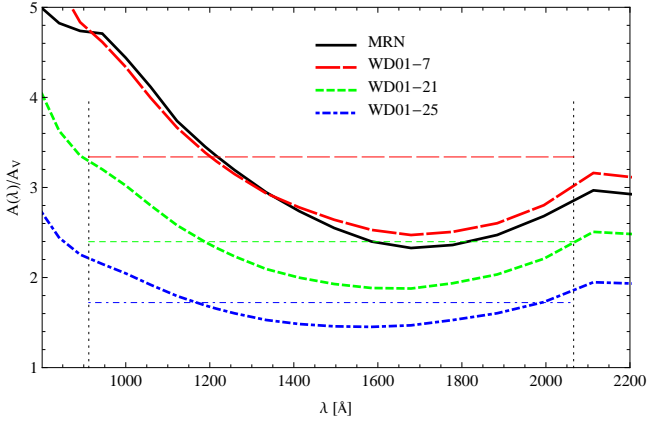}}
\caption{Relative extinction for the dust models of MRN (solid) , WD01-7 (long-dashed; red in electronic version); WD01-21 (short-dashed; green in the electronic version) and WD01-25 (dot-dot-dashed; blue in the electronic version) of interstellar dust. The vertical dotted lines at wavelengths  of 912 and 2066 {\AA}, corresponding to 13.6 and 6 eV, show the range of averaging. The horizontal lines show $k_{\rm FUV}$ for the considered models.}\label{relext}
\end{figure}

Fig. \ref{relext} shows $A_\lambda/A_{\rm V}$ for the dust models tested in this paper: MRN (solid), WD01-7 (long-dashed; red in the electronic version), WD01-21 (short-dashed;  green in the electronic version), and WD01-25 (dot-dot-dashed; blue in the electronic version). The FUV range relevant for photoelectric heating and photo-dissociation reactions extends between 13.6 and 6 eV\footnote{H$_2$ only absorbs FUV photons via Lyman and Werner electronic transitions in the 912-1100\,\AA\ range.}. Consequently, we define an FUV-to-V color as $k_{\rm FUV}$\,= 
$\langle A(\lambda)/A(V)\rangle_\lambda$ \,$\equiv \langle A_\lambda/A_{\rm V}\rangle_\lambda$ where the averaging is performed over an energy from 6 to 13.6 eV and  $A_{\rm V}$ is the visual extinction. 
The vertical dotted lines in Fig.~\ref{relext} indicate this range of averaging and the horizontal lines show $k_{\rm FUV}$ for the considered models. Note that the MRN and WD01-7 dust models have the same value of $k_{\rm FUV}=3.339$. All values of the average relative dust extinction are collected in Tab. \ref{table:1}. 
From Fig. \ref{relext} it can be seen that $k_{\rm FUV}$ is significantly decreased in the WD01-21 and WD01-25 dust models. This is consistent with the higher values of $R_V$, i.e., a shallower gradient of $A(\lambda)/A_V$ toward shorter wavelengths causing a lower UV extinction for the same $A_V$ (see Fig. \ref{relext}). This is the main source for the different UV attenuation properties of different dust types.

\subsubsection{Influence of radiative transfer and scattering\label{influencert}}

\begin{figure}
\resizebox{\hsize}{!}{\includegraphics{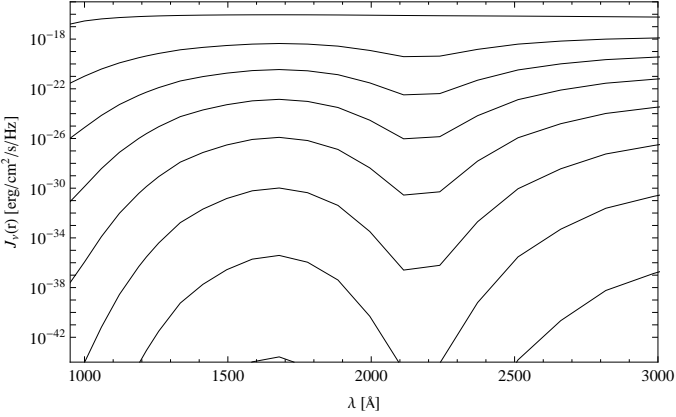}}
 \caption{Mean intensity $J_\lambda(r)$ for a model clump of $M=$10 \msol, $n=10^5$~cm$^{-3}$, and $\chi=1000$. The lines show $J_\lambda(r)$ for radii in steps of 10\% of $R_\mathrm{tot}$. }
 \label{jnur2lambda}
 \end{figure}

As a result of the new radiative transfer computations we obtain the full, wavelength-dependent FUV radiation field  $J_\lambda(r)=\frac{1}{4\pi}\int I_\lambda d\Omega$ at all clump radii, where $I_\lambda$ is the specific intensity at wavelength $\lambda$. In Fig. \ref{jnur2lambda} we plot $J_\lambda(r)$ for radii in steps of $R_\mathrm{tot}/10$ for the WD01-7 model. The effect of the prominent 2200 \AA\ ``bump'' is plainly visible in the enhanced reduction of $J_\lambda$ with decreasing radius. 

The FUV optical depth 
\begin{equation}
\tau_{\rm FUV} = k_{\rm FUV} \tau_{\rm V} = N_{\rm H} \sigma_{\rm D,FUV}
\end{equation}
can be described by the effective dust extinction cross section
\begin{equation}\label{ktausigmadust}
\sigma_{\rm D,FUV} = k_{\rm FUV} \, m_{\rm H} \, \Psi \, \langle{{K}_{\rm ext}({\rm V}})\rangle\;.
\end{equation}
For adopted dust models, the obtained values of $\sigma_{\rm D,FUV}$ are collected in Tab.~\ref{table:1}. 

The original KOSMA-$\tau$ code did not include any angular dependence
of the dust scattering but implicitly assumed pure forward scattering 
\citep{SD89} in terms of an extinction coefficient $\sigma_{\rm D,FUV}$.
As long as the models use the same value of $\sigma_{\rm D,FUV}$,
the radiative transfer should always give the same scaling of
the UV intensity with exp(\,-$N_{\rm H} \sigma_{\rm D,FUV}$\,)
at high optical depths because the radiation from scattering 
at other angles is quickly damped because of the longer optical
path, but the scattering will increase the intensity close to 
the surface \citep{FlanneryRoberge1980}.

To study the influence of FUV scattering we performed a set of model calculations with different scattering properties but the same effective value of $\sigma_{D,FUV}=1.76\times 10^{-21}$~cm$^2$:
\begin{enumerate}
\item The old \ktau\ FUV calculations using pure forward scattering. 
\item The MCDRT result for a MRN dust distribution providing the same $\sigma_{D,FUV}$, but where the albedo $\omega$ is artificially set to zero.
\item The MCDRT result for the MRN distribution with a full treatment of scattering and absorption. 
\end{enumerate}
In Appendix~\ref{scattering} we discuss how the assumption of isotropic scattering compares to more realistic scattering properties and demonstrate that it poses a clear improvement compared to the pure forward scattering case used in our previous model.

\begin{figure}
\resizebox{\hsize}{!}{\includegraphics{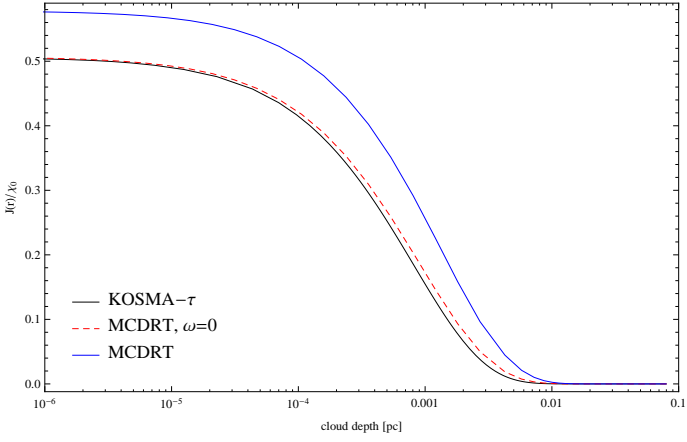}}
\caption{Mean intensity of the UV radiation vs. cloud depth. The black line corresponds to the old \ktau\  result for pure absorption. The blue line corresponds to the MCDRT result for an MRN distribution with full treatment of scattering and absorption. The dashed line is the result for an MRN distribution with the albedo artificially set to zero.}
\label{FUV2depth1}
\end{figure}

Figure \ref{FUV2depth1} shows a comparison of the depth-dependant mean FUV intensity scaled by the unattenuated mean FUV intensity $J(r)/\chi_0$ for these three models. $J(r)$ is the total mean intensity, i.e., averaged over the full solid angle and integrated over the full wavelength range:  $J(r)=\int_\mathrm{912\AA}^\mathrm{2066\AA}J_\lambda(r)d\lambda.$ $\chi_0$ is the corresponding total mean intensity of the FUV field in the absence of the molecular cloud, i.e., the full unattenuated photon flux coming from $4\pi$ solid angle. For $r=R_\mathrm{tot}$, models 1) and 2) are slightly larger than 0.5, which would be the theoretical value for a fully opaque and very large clump. Higher values indicate the contribution from angles slightly larger than 90\deg. By contrast model 3 shows a value of 0.58, indicating significant contributions to $J(r)$ by scattered photons. Consequently, the mean intensity is higher for model 3) throughout the whole clump because of a large amount of scattered photons. The small differences between models 1) and 2) result from the full wavelength treatment compared to the attenuation by an average $\tau_{\rm FUV}$ only.
For high optical depths models 2) and 3) converge to the same intensities as all sidewards scattered photons turn insignificant there because of the longer optical path \citep{FlanneryRoberge1980}.

Eq. (\ref{ktausigmadust})  
turns out to be a very good approximation when only 
forward-scaterring occurs, which is clearly reproduced by the detailed 
radiative transfer computations with the albedo set to zero.
However, only the full radiative transfer treatment can produce the full spatial and spectral knowledge of the radiation field and the proper accounting for scattered photons.

\begin{figure}
\resizebox{\hsize}{!}{\includegraphics{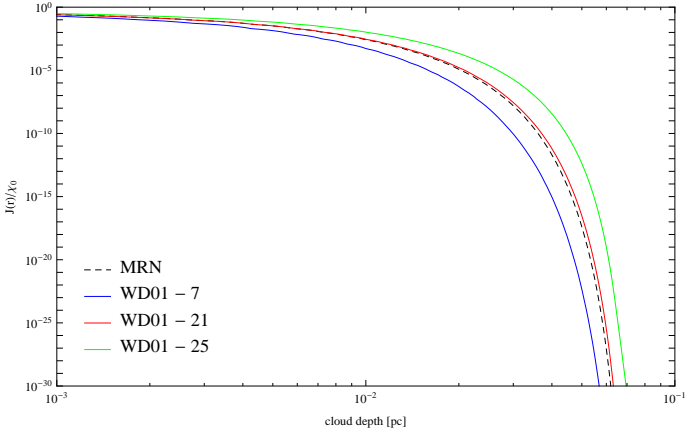}}
\caption{Mean intensity of the UV radiation $J(r)$ scaled to the unshielded (empty space) Draine field $\chi_0$ vs. cloud depth. Different lines denote different dust compositions.}

\label{FUV2depth2}
\end{figure}
 
The influence of dust composition on the resulting depth-dependent FUV intensity in the clump can be seen in Fig. \ref{FUV2depth2}. For the FUV intensity in the WD01 models, the following is true throughout the whole clump: WD01-7 < WD01-21 < WD01-25.

\subsubsection{Influence on photo-reactions\label{sect:influencephotoreactions}}
The local UV intensity acts on the chemistry of the clump by heating the gas and dust and by inducing ionization and dissociation reactions. This photo-process $j$ with a wavelength-dependent cross section $\sigma_j(\lambda)$ proceeds at a rate \citep{roberge1991} 
\begin{equation}\label{photo1}
\Gamma_j(r)=4\pi\int_{\lambda_H}^{\lambda_j}J_\lambda(r)\sigma_j(\lambda)d\lambda .
\end{equation}
$J_\lambda(r)$ is the mean intensity of radiation at radius $r$ in photons cm$^{-2}$s$^{-1}$ nm$^{-1}$. The integral runs from $\lambda_H=912$~\AA\ to the threshold wavelength $\lambda_j$ for the process $j$\footnote{A database of photo-ionization and photo-dissociation cross sections for common astrophysical species can be found at  http://home.strw.leidenuniv.nl/$\sim$ewine/photo/. For more details see \citet{vdishoeck2006} and \citet{vhemert2008}}. Astrochemical databases often provide parametrized reaction rate coefficients to allow the calculation of these photo-reaction rates. For instance, UDfA gives the rate coefficients in the form
\begin{equation}\label{photo2}
\Gamma_j(A_V)=\chi_0 \alpha_j\exp(-\gamma_j A_V)
\end{equation}
assuming an attenuation of the radiation 
 provided by a standard MRN dust model in a plane-parallel configuration using the radiation transfer results from \citet{FlanneryRoberge1980} parametrized in terms of the perpendicular visual extinction $A_V$. $\chi_0$ is the FUV field strength at the edge of the cloud and the unshielded, free-space rate coefficient $\alpha_j$ assumes a mean FUV intensity of unity (in units of the Draine field) and FUV photons coming from all directions. At the edge of an optically thick molecular cloud the rate coefficient is about half of this value, depending on the dust scattering properties.

\begin{figure}
\resizebox{\hsize}{!}{\includegraphics{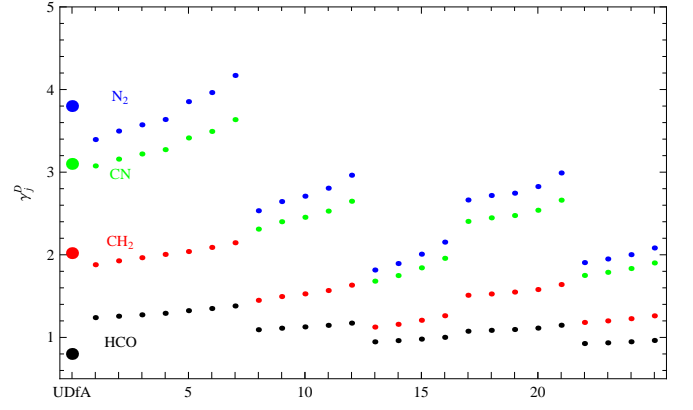}}
\caption{Fit results for $\gamma_{j,fit}^D$ for \ce{HCO} (black), \ce{CH2} (red), \ce{CN} (green), and \ce{N2} (blue) for all dust models from \citet[][ their Tab.~1]{WD01}. The abscissa gives the number of the dust model. The large points show the original values of $\gamma_j$ for the respective molecules.}
\label{gammaOrder}
\end{figure}

\begin{figure}
\resizebox{\hsize}{!}{\includegraphics{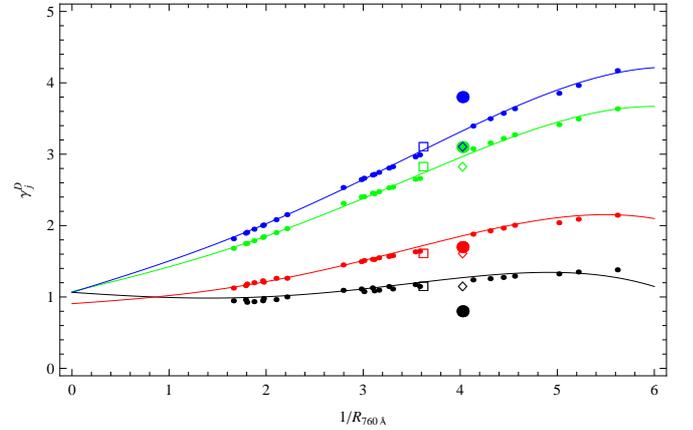}}
\caption{Comparison of explicitly fitted $\gamma_j^D$ for \ce{HCO} (black), \ce{CH2} (red), \ce{CN} (green), and \ce{N2} (blue) for all dust models from \citet[][ their Tab.~1]{WD01} with scaling relationship Eq.~\ref{gammascaling}. The large points show the original values of $\gamma_j$ from UDfA, the open diamonds show the values calculated using an MRN dust model, and the open squares denote the explicitly calculated values assuming $1/R_{760\AA}^\mathrm{MRN}=3.62$. }
\label{newgamma}
\end{figure}

The unshielded rate coefficient $\alpha_j$ does not depend on the dust properties and we concentrated on the attenuation properties, parametrized in form of the $\gamma_{j}^D$. Changing either the dust properties, i.e., the wavelength-dependant FUV attenuation, or the spectral distribution of the FUV field leads to a different result for $\Gamma_j$ and consequently to a different fit parameter $\gamma_j$. However, performing the wavelength-dependent integration of Eq.~\ref{photo1} for every reaction and every radial point is often beyond the scope of  PDR models, e.g., because it is too time-consuming or because neither dust properties nor ionization cross sections $\sigma_j(\lambda)$ are known accurately enough to justify the computational effort of a full integration of the wavelength-dependent radiative transfer. Furthermore, knowledge of the wavelength-dependent $\sigma_j$ is not available for all astrophysically relevant species and models have to apply Eq.~\ref{photo2} even if they can perform the full integration in principle.
 
Here we derive a scaling relation 
between the $\gamma_j$ values provided by a data base like UDfA and corresponding fit values for a particular dust model 
$\gamma_j\rightarrow \gamma_j^D$ to calculate the rate coefficient $\Gamma_j^D(A_\mathrm{V})$ for species $j$ and dust sort $D$. 
For a direct comparison with  UDfA values we started from a model cloud with a mass of $10^3$~M$_\odot$ and a radius of 1.71~pc, large enough to neglect the effects of the spherical symmetry close to the surface, isotropically illuminated and calculated the photo-dissociation and photo-ionization rates of 72 species provided by \citet{vdishoeck2006,vhemert2008} for all 25 dust models $D$ from \citet{WD01} according to Eq.~\ref{photo1}.

 To this end, we performed least-squares fits to $\Gamma_j(A_V)/\Gamma_j(0)=\exp(-\gamma_{j,fit}^D A_\mathrm{V})$ up to an $A_\mathrm{V}=10$ to determine new $\gamma_{j,fit}^D$ coefficients for all $D$ and $j$. 
Figure~\ref{gammaOrder} shows the new  $\gamma_{j,fit}^D$ for four example molecules, \ce{HCO}, \ce{CH2}, \ce{CN}, and \ce{N2} for all 25 dust models. 
 
The big dots on the left show the original value $\gamma_j$ from UDfA\footnote{\citet{vdishoeck2006} and \citet{vhemert2008} provide their own fitted values of $\gamma_{j,\mathrm{EvD}}$ (up to an maximum $A_\mathrm{V}=3$) for all species for which they also give tables of $\sigma_j(\lambda)$. 
}. The figure shows that the new $\gamma_{j,fit}^D$ show the same behavior along the abscissa for all four example species. There are also three branches of $\gamma_{j,fit}^D$, visible as prominent quantitative steps in Fig.~\ref{gammaOrder} that correspond to the respective $R_\mathrm{V}$ values. 

The next step is to identify what dust model property is best correlated with $\gamma_{j,fit}^D$. An extensive analysis of the dust properties reveals that we can find a wavelength $\lambda=760\AA$ for which 
\begin{equation}\label{rinv}
1/R_\lambda=\left(\frac{A(V)}{A(\lambda)-A(V)}\right)^{-1}     
\end{equation}
shows a monotonic ordering for the $\gamma_{j,fit}^D$  of the 25 dust models. Consequently, we chose $1/R_{760\AA}^D$ as the parameter describing the dust. Table~\ref{table:rinv} gives  $1/R_{760\AA}^D$ for all 25 WD01 dust models. 
\begin{table}[htb]
\begin{center}
\caption{Dust model parametrization of all WD01 dust models. See \citet{WD01} for more details.}
\label{table:rinv}
\begin{tabular}    {r c c c c r }
\hline\hline
$D$\tablefootmark{a}&$R_\mathrm{V}$\tablefootmark{b}&$10^5b_\mathrm{c}$\tablefootmark{c}&Case\tablefootmark{d}&$1/R_{760\AA}^D$\tablefootmark{e}&order\tablefootmark{f}\\ \hline
1 & 3.1 & 0.0 & A & 4.13  & 19 \\
2 & 3.1 & 1.0 & A & 4.31  & 20 \\
3 & 3.1 & 2.0 & A & 4.44  & 21 \\
4 & 3.1 & 3.0 & A & 4.56  & 22 \\
5 & 3.1 & 4.0 & A & 5.02  & 23 \\
6 & 3.1 & 5.0 & A & 5.22  & 24 \\
7 & 3.1 & 6.0 & A & 5.62  & 25 \\
8 & 4.0 & 0.0 & A & 2.79  & 9 \\
9 & 4.0 & 1.0 & A & 2.98  & 10 \\
10 & 4.0 & 2.0 & A & 3.10  & 12 \\
11 & 4.0 & 3.0 & A & 3.26  & 15 \\
12 & 4.0 & 4.0 & A & 3.54  & 17 \\
13 & 5.5 & 0.0 & A & 1.66  & 1 \\
14 & 5.5 & 1.0 & A & 1.79  & 2 \\
15 & 5.5 & 2.0 & A & 1.97  & 6 \\
16 & 5.5 & 3.0 & A & 2.21  & 8 \\
17 & 4.0 & 0.0 & B & 3.01  & 11 \\
18 & 4.0 & 1.0 & B & 3.11  & 13 \\
19 & 4.0 & 2.0 & B & 3.16  & 14 \\
20 & 4.0 & 3.0 & B & 3.30  & 16 \\
21 & 4.0 & 4.0 & B & 3.58  & 18 \\
22 & 5.5 & 0.0 & B & 1.80  & 3 \\
23 & 5.5 & 1.0 & B & 1.88  & 4 \\
24 & 5.5 & 2.0 & B & 1.96  & 5 \\
25 & 5.5 & 3.0 & B & 2.10  & 7 \\\hline
\end{tabular}
\end{center}
\tablefoot{\\
\tablefoottext{a}{Dust model $D$, see Tab.~1 in WD01}\\
\tablefoottext{b}{$R_\mathrm{V}=A(V)/E_\mathrm{B-V}$, ratio of visual extinction to reddening.}\\
\tablefoottext{c}{C abundance in double log-normal very small grain population.}\\
\tablefoottext{d}{Case A: variable grain volumes. Case B: grain volumes fixed at approximately the values found for $R_\mathrm{V}=3.1$.}\\
\tablefoottext{e}{$1/R_\lambda=\left(A(V)/(A(\lambda)-A(V))\right)^{-1}$}\\
\tablefoottext{f}{Ordering of the various dust models from lowest to highest value of $1/R_{760\AA}$.}
}
\end{table}
Using this parameter, we can approximate the dependence of $\gamma_j^D$ on $\gamma_j$ and the dust model by
\begin{eqnarray}\label{gammascaling}
\gamma_j^D&=& w_1+ w_2 \left(\frac{1}{R_{760 \text{\AA}}^D}\right) + w_3 \left(\frac{1}{R_{760 \text{\AA}}^D}\right)^3 + w_4 \left(\frac{1}{R_{760
   \text{\AA}}^D}\right)^4 \\ && 
+ w_5 \gamma_j + w_6 \gamma_j^3 + w_7 \gamma_j^4 + w_8 \frac{\gamma_j}{R_{760
   \text{\AA}}^D} + w_9 \frac{\gamma_j^2}{R_{760   \text{\AA}}^D}   .\nonumber
\end{eqnarray}

 \begin{table}[htb]
 \begin{center}
 \caption{Fit parameters in Eq.~\ref{gammascaling}.}
 \label{table:gammascaling}
 \begin{tabular}{l l}
 \hline\hline
 \multicolumn{2}{c}{fit parameter}\\ \hline
 $w_1$ & 1.39460 \\
 $w_2$ & -0.27655 \\
 $w_3$ & 0.02053 \\
 $w_4$ & -0.00295 \\
 $w_5$ & -0.45990 \\
 $w_6$ & 0.08725 \\
 $w_7$ & -0.01616 \\
 $w_8$ & 0.24747 \\
 $w_9$ & -0.01677 \\\hline
 \end{tabular}
 \end{center}
 \end{table}

Equation~\ref{gammascaling} reproduces the explicitly calculated attenuation of the rate coefficients within an accuracy of $\sigma=0.05$, and a maximum absolute residual of 0.22. This is comparable to the uncertainties of $\gamma_{j,fit}^D$ due to fitting $\Gamma_j(A_V)/\Gamma_j(0)$ up to a different maximum $A_\mathrm{V}$.  
Note that for the fitting we removed seven species (photo-dissociation of
\ce{CH+}, \ce{SH+}, \ce{OH+}, \ce{HCO+}, \ce{CO}, \ce{O2+}, and \ce{SiO}
) from the data set because their $\gamma_j^D$ represent strong outliers with respect to the remainder of the data set.
We discuss possible reasons for these deviations in Appendix~\ref{appendix:rescaling}. 
To apply Eq.~\ref{gammascaling} to them, we had to use corrected coefficients $\hat{\gamma}_j$ instead of the UDfA vales when computing $\gamma_j^{D}$.
The  $\hat{\gamma}_j$ values are given in Tab.~\ref{table:gammaoutliers}.

 \begin{table}[htb]
 \begin{center}
 \caption{Corrected $\hat{\gamma}_j$ for the outlying species. }
 \label{table:gammaoutliers}
 \begin{tabular}{l l l}
 \hline\hline
 species&$\gamma_j$& $\hat{\gamma}_j$ \\ \hline
 \ce{CH+} & 2.5 & 2.11 \\
 \ce{SH+} & 1.8 & 1.39  \\
 \ce{OH+} & 1.8 & 2.76  \\
 \ce{HCO+} & 2. & 3.036 \\
 \ce{CO} & 2.5 & 3.03 \\
 \ce{O2+} & 2. & 1.61 \\
 \ce{SiO} & 2.3 & 1.87  \\
 \hline
 \end{tabular}
 \end{center}
 \end{table}

It turns out that the original MRN distribution with  $1/R_{760\AA}^\mathrm{MRN}=4.03$ does not exactly follow the parametrization of Eq.~\ref{gammascaling}, based on the WD01 dust models, but that it can be easily used in the same function when adjusting the parameter $1/R_{760\AA}^\mathrm{MRN}=3.62$.
 Figure~\ref{newgamma} compares  $\gamma_{j,fit}^D$ for \ce{HCO}, \ce{CH2}, \ce{CN}, and \ce{N2} with the results of Eq.~\ref{gammascaling}.
In addition we show the MRN results in the plot: the original $\gamma_j$, the explicitly calculated $\gamma_j^\mathrm{MRN}$, and the  $\gamma_j^\mathrm{MRN}$ assuming $1/R_{760\AA}^\mathrm{MRN}=3.62$ using  large points, open diamonds, and open squares, respectively (the arrows in the figure demonstrate the shifting of the rescaled $\gamma_j^D$ closer to Eq.~\ref{gammascaling}).

It is unclear why the $\gamma_j$ from UDfA, i.e., the MRN values, do not follow the parametrization of Eq.~\ref{gammascaling}. There is a number of possible reasons for the deviation. \citet{udfa06} did not use a Draine FUV radiation field but assumed a 10000~K black-body radiation field. The different spectral shape usually accounts for approximately 10\% difference but can in a few special cases result in a significant difference. Secondly, different $A_\mathrm{V,max}$ up to which the $\gamma_j$ is fitted give different values of $\gamma_j$( see Appendix~\ref{appendix:rescaling} for a more detailed discussion). The details of the radiation scattering is another possible reason for different results. Assuming different anisotropy factors $g$ and dust albedo $\omega$ will change the radiation field and accordingly affect the $\gamma_j$ (see Appendix~\ref{scattering} for a discussion on the effect of different values of $g$).

We conclude that the scaling relation Eq.~\ref{gammascaling} is able to rescale the tabulated $\gamma_j$ from UDfA to the dust-specific $\gamma_{j,fit}^D$ (fitted to the fully calculated $\Gamma_j(A_V)/\Gamma_j(0)$) with an accuracy of about 10\%.

\subsection{Dust temperatures\label{sect:dusttemp}}

\begin{figure}
\resizebox{\hsize}{!}{\includegraphics{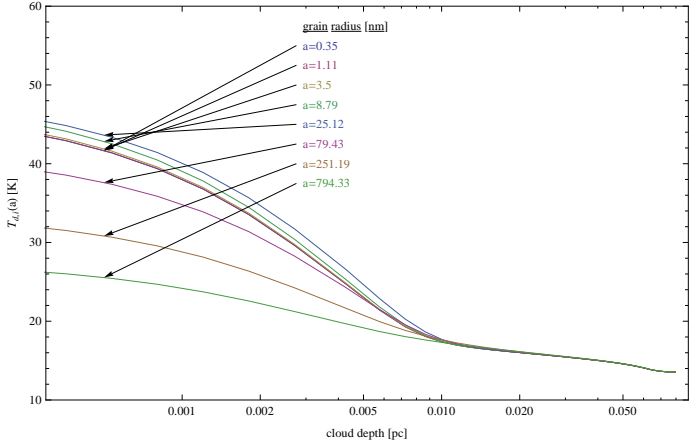}}
\caption{ Dust temperatures of silicate grains for a model clump of $M=$10 \msol, $n=10^5$~cm$^{-3}$, and $\chi=1000$ using the WD01-7 dust model. The different lines denote different grain sizes. }
\label{TDWD07sili}
\end{figure}

\begin{figure}
\resizebox{\hsize}{!}{\includegraphics{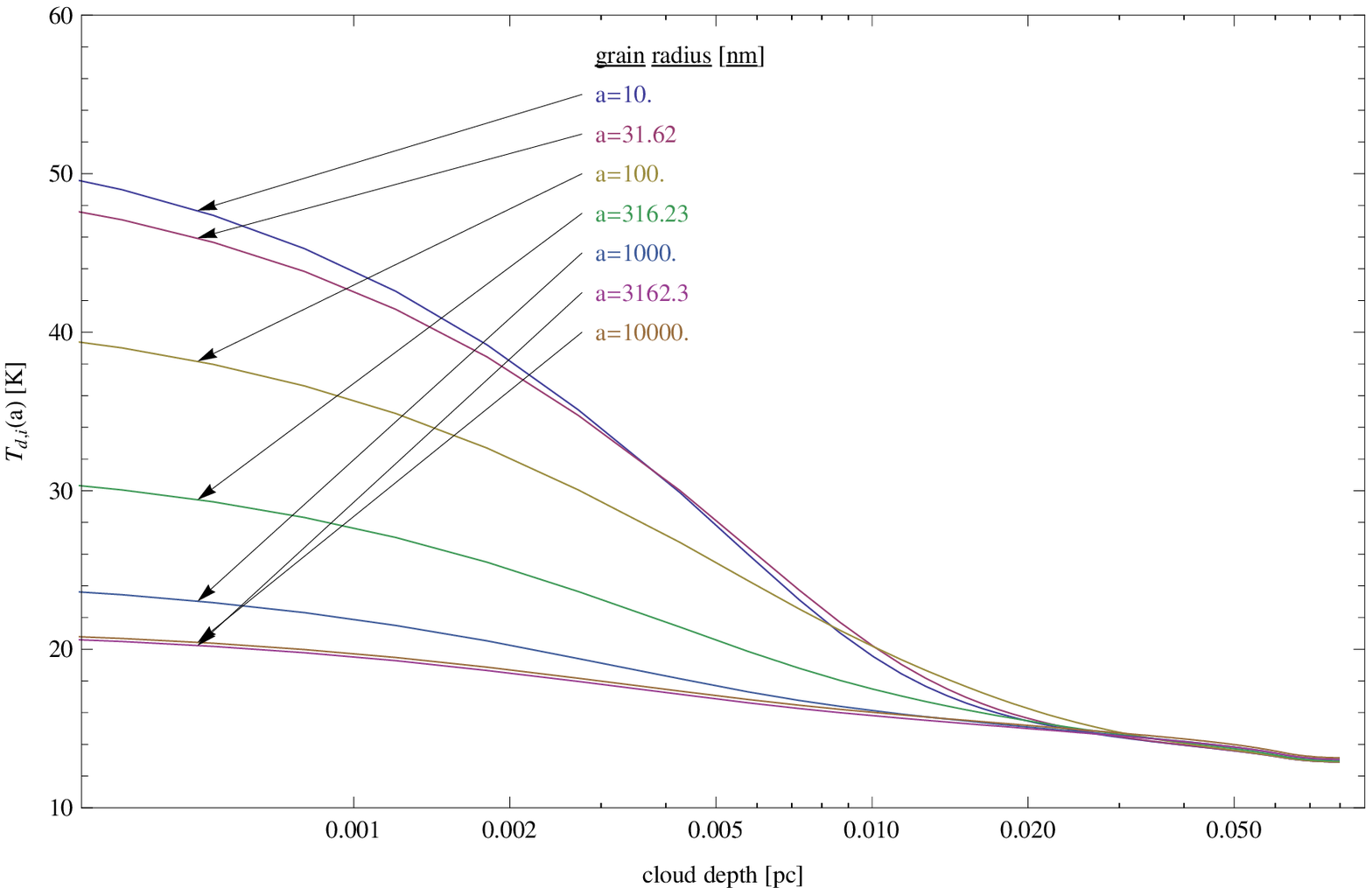}}
\caption{Dust temperatures of carbonaceous grains for a model clump of $M=$10 \msol, $n=10^5$~cm$^{-3}$, and $\chi=1000$ using the WD01-7 dust model. The different lines denote different grain sizes. }
\label{TDWD07graph}
\end{figure}

\begin{figure}
\resizebox{\hsize}{!}{\includegraphics{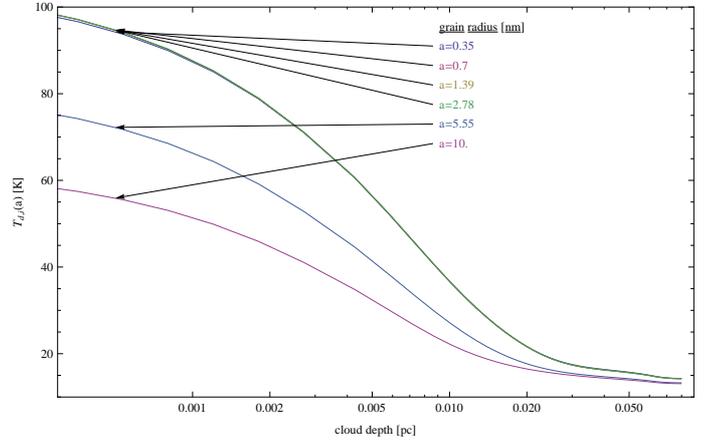}}
\caption{Dust temperatures of neutral PAHs for a model clump of $M=$10 \msol, $n=10^5$~cm$^{-3}$, and $\chi=1000$ using the WD01-7 dust model. The different lines denote different PAH sizes. }
\label{TDWD07nPAH}
\end{figure}

Dust temperatures are calculated independently for each dust component and  size.
Figures \ref{TDWD07sili}, \ref{TDWD07graph}, and \ref{TDWD07nPAH} show depth-dependent grain temperatures for a number of differently sized grains for three of the four components. The plot for the ionized PAHs is omitted because their temperature behaves like that of neutral PAHs.  The general behavior is that smaller grains exhibit higher grain temperatures.

\begin{figure}
\resizebox{\hsize}{!}{\includegraphics{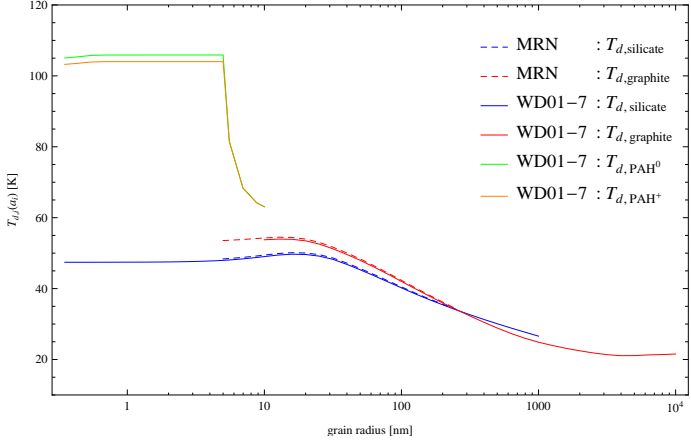}}
\caption{ Grain temperature for the components of the MRN and WD01-07 dust models as a function of the radius $a_i$ in nm. The grain temperatures are calculated at the surface of a model clump of $M=$10 \msol, $n=10^5$~cm$^{-3}$, and $\chi=1000$. The different lines denote different dust components. Dashed lines show the MRN components and solid lines show the WD01-07 components. }
\label{TDvsradius}
\end{figure}

Figure \ref{TDvsradius} compares grain temperatures of the components of the MRN and WD01-07 dust model at the surface of a model clump as a function of grain radius $a_i$.  Graphites tend to have a higher temperature than silicate grains except for very large grain sizes. Small grains tend to be hotter than bigger grains. The PAHs are heated much more efficiently than the larger grains and have significantly higher grain temperatures. For simplicity, we deliberately neglected the non-equilibrium heating of very small grains and PAHs here and assumed equilibrium heating in this paper. In a subsequent paper, we will include the comparison with real observations accounting for the stochastic heating of the very small particles.  
The dependence on $a_i$ vanishes for sufficiently high extinctions i.e., once most of the remaining photons have wavelengths longer than the grains. The resulting equilibrium grain temperature then only depends on the absorption coefficient of the material.

The detailed knowledge of the dust temperature as a function of radius, grain size, and grain type enables us to treat all dust-related processes separately for all grain sizes and types. 
However, since most relevant processes are grain surface reactions, e.g., formation of the \HH\ or photoelectric heating, it is usually sufficient to calculate surface-averaged quantities. For the $i$-th dust component we can define the relevant mean dust temperature as
\begin{equation}\label{Tdaver1}
\langle T_{d,i}\rangle_a=\frac{\int_{a_{i,-}}^{a_{i,+}}T_{d,i}(a)n_i(a) a^2 da}{\int_{a_{i,-}}^{a_{i,+}}n_i(a)a^2 da} .
\end{equation}
Averaging over all dust sorts is done according to
\begin{equation}\label{Tdaver2}
\langle T_{d}\rangle_\mathrm{NDS}=\sum_{i=1}^\mathrm{NDS}\left\langle T_{d,i}\right\rangle_a \frac{\sigma_{d,i}}{\sigma_d}\, 
\end{equation}
with the geometrical cross section of the $i$-th sort $\sigma_{d,i}$ and the total dust cross section  $\sigma_d$. We will use the short-hand notation $T_d=\langle T_{d}\rangle_\mathrm{NDS}$ and $T_{d,i}=\langle T_{d,i}\rangle_a$ from here on.

\begin{figure}
\resizebox{\hsize}{!}{\includegraphics{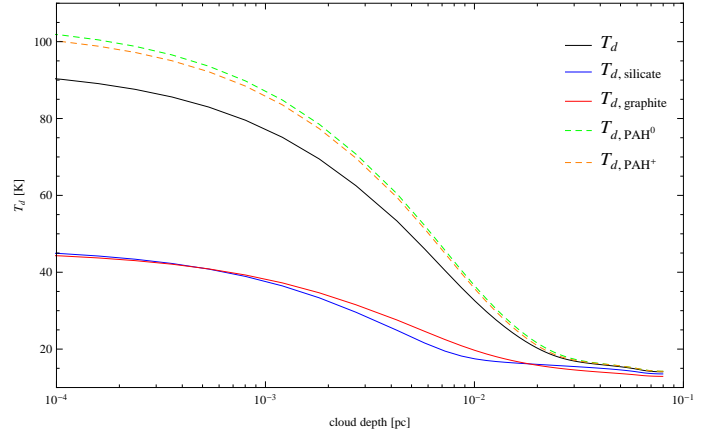}}
\caption{Mean dust temperature for the WD01-7 dust model for a model clump of $M=$10 \msol, $n=10^5$~cm$^{-3}$, and $\chi=1000$. The different lines denote the average temperature of the individual dust components. The black line shows the overall mean dust temperature. }
\label{Tdust2}
\end{figure}

Figure \ref{Tdust2} shows the mean dust temperature $T_{d,i}$ for the WD-7 dust model as a function of cloud depth for all four dust components. As at the surface (Fig.~\ref{TDvsradius}), silicates, indicated by the solid blue line, are cooler than graphite grains (solid, red line). 
Graphite and silicate grain temperatures are very close, but the mean silicate temperature drops slightly faster for increased cloud depths than the graphites.
The PAHs have a much higher  $T_{d,i}$. Neutral PAHs tend to be marginally hotter than the ionized PAHs. The black line denotes the average dust temperature $T_d$ for the WD01-7 dust. 

\begin{figure}
\resizebox{\hsize}{!}{\includegraphics{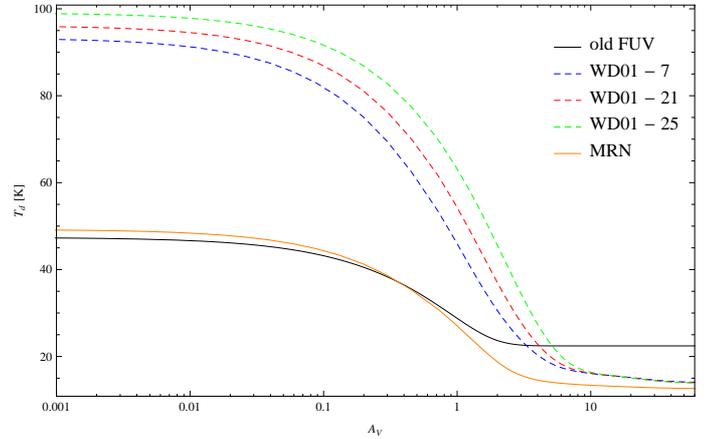}}
\caption{Mean dust temperature as a function of \Av\  for a model clump of $M=$10 \msol, $n=10^5$~cm$^{-3}$, and $\chi=1000$. The different lines denote different dust compositions. }
\label{Tdust}
\end{figure}

The influence of the different dust models can be seen from Fig. \ref{Tdust}. The black line shows the old dust temperature calculation in \ktau\ following \citet{HTT91}, which gives a central dust temperatures $T_d \approx 22$~K. Accounting for the full wavelength dependence and the detailed grain size distribution gives a significantly lower $T_d\approx 13$~K at the center of the model clump. On the other hand, it also leads to warmer dust at the surface of the cloud compared to \citet{HTT91}.

The dust temperature for the WD01 models follow their corresponding FUV intensities (see Fig. \ref{FUV2depth2}). The MRN model has a much lower $T_d$. Figure \ref{TDvsradius} shows that the $T_{d,i}$ for MRN and WD01-7 are very similar for equal grain sizes. However, the temperature of the PAHs in the WD01-7 model contributes significantly to the average dust temperatures in Eq. (\ref{Tdaver2}) and leads to higher $T_d$ in models with PAH content.

The lower central dust temperatures of models with WD01 dust types also influences the gas temperatures. Models with WD01 dust show  significantly lower central gas temperatures because gas-grain collisions are becoming the dominant cooling process. 
  
\subsection{\HH\ formation\label{sect:h2formation}}

\begin{table*}[htb]
\begin{center}
\caption{Parameter of the WD01 and MRN dust models}
\label{table:1}
\begin{tabular}    {l l  l l l  l  l }
\hline\hline
model&parameter&silicates&graphite&PAH$^0$&PAH$^+$&total\\ \hline

MRN&$a_-$ [\AA]&50&50&&&\\
& $a_+ [\mu\mathrm{m}]$&0.25&0.25&&&\\
&$k_{\rm FUV}$&&&&&3.339 \\
&$\sigma_{\rm D,FUV}\,[cm^2]$&$5.43 \times 10^{-22}$&$1.22 \times 10^{-21}$&&&$1.76 \times 10^{-21}$\\
&$\sigma_d$\,[cm$^2$]&$6.05\times 10^{-22}$&$5.65\times 10^{-22}$&&&$1.17 \times 10^{-21}$\\
&$\Psi$ [\%]&0.598&0.357&&&0.955\\

\hline

WD01-07&$a_-\,$ [\AA]&3.5&100&3.5&3.5&\\
& $a_+\, [\mu\mathrm{m}]$&1.&10.&0.01&0.01&\\
&$k_{\rm FUV}$&&&&&3.339 \\
&$\sigma_{\rm D,FUV}\,[cm^2]$&$9.59 \times 10^{-22}$&$6.94 \times 10^{-22}$&$4.73 \times 10^{-23}$&$4.58 \times 10^{-23}$&$1.75 \times 10^{-21}$\\
&$\sigma_d$\,[cm$^2$]&$1.14\times 10^{-21}$&$1.34\times 10^{-22}$&$2.71\times 10^{-21}$&$2.71\times 10^{-22}$&$6.69 \times 10^{-21}$\\
&$\Psi$ [\%]&0.828&0.0.228&0.038&0.038&1.130\\
\hline

WD01-21&$a_-\,$ [\AA]&3.5&100&3.5&3.5&\\
& $a_+\, [\mu\mathrm{m}]$&1.&10.&0.01&0.01&\\
&$k_{\rm FUV}$&&&&&2.398 \\
&$\sigma_{\rm D,FUV}\,[cm^2]$&$8.56 \times 10^{-22}$&$3.06 \times 10^{-22}$&$2.44 \times 10^{-23}$&$2.37 \times 10^{-23}$&$1.21 \times 10^{-21}$\\
&$\sigma_d$\,[cm$^2$]&$4.93\times 10^{-22}$&$9.99\times 10^{-23}$&$1.85\times 10^{-21}$&$1.85\times 10^{-21}$&$4.29 \times 10^{-21}$\\
&$\Psi$ [\%]&0.815&0.0.252&0.027&0.027&1.120\\
\hline

WD01-25&$a_- \,$[\AA]&3.5&100&3.5&3.5&\\
& $a_+\, [\mu\mathrm{m}]$&1.&10.&0.01&0.01&\\
&$k_{\rm FUV}$&&&&&1.722 \\
&$\sigma_{\rm D,FUV}\,[cm^2]$&$6.78 \times 10^{-22}$&$1.40 \times 10^{-22}$&$1.22 \times 10^{-23}$&$1.18 \times 10^{-23}$&$8.41 \times 10^{-22}$\\
&$\sigma_d$\,[cm$^2$]&$2.12\times 10^{-22}$&$5.08\times 10^{-23}$&$1.36\times 10^{-21}$&$1.36\times 10^{-21}$&$2.98 \times 10^{-21}$\\
&$\Psi$ [\%]&0.812&0.0.268&0.019&0.019&1.118\\
\hline
\end{tabular}
\end{center}
\end{table*}

Molecular hydrogen is the most abundant molecule in the universe. The formation of \HH\  most effectively takes place on the surface of dust grains. The proper treatment of the \HH\ formation is crucial in any numerical PDR model \citep[see e.g.][ for a recent update of the Meudon PDR code.]{lebourlot2012}. We updated the \ktau\ calculation of the \HH\ formation efficiency $R_{\mathrm{H}_2}$ from the method described by SD89 to account for physisorption and chemisorption binding energies on silicate and graphite surfaces \citep[][(hereafter CT)]{cazaux2002,cazaux2004,cazaux2010erratum}. 

 This model is based on two main points: (1) H atoms can stick to the grain surface in two different binding sites, a physisorption site and a chemisorption site. Physisorption is a weak atom-surface binding resulting from a Van der Waals force (dipole-dipole interaction). Chemisorption is a strong atom-surface binding interaction involving the valence electrons, known as covalent bond. These interactions determine whether atoms move on the surface, and thus, how quickly they can meet a recombination partner. (2) The mobility of a surface H atom results from the combination of two physical processes: tunneling and thermal diffusion; tunneling dominates at the lowest temperatures while thermal diffusion is most important at the highest temperatures. The calculation of the transmission coefficients, to go from site to site, are given in \citet{cazaux2004,cazaux2010erratum}.

At low temperature ($T\le100$ K), \HH\ formation involves the migration of physisorbed H atoms. At
higher temperatures ($T\ge 100$ K), \HH\ formation results from
chemisorbed H recombination. The presence of these two types
of binding sites allows \HH formation to proceed relatively efficiently at low and elevated temperatures.

For the dust component $i$, the formation rate can be written as
\begin{equation}
R_{\mathrm{H}_2,i}=\frac{1}{2}n_\mathrm{H} v_\mathrm{H} n_{d,i} \sigma_{d,i}\epsilon_{\mathrm{H}_2,i}S_\mathrm{H} , \label{h2rate}
\end{equation}
where $n_\mathrm{H}$ and $v_\mathrm{H}$ are the number density and thermal velocity of H atoms in the gas phase, $n_{d,i} \sigma_{d,i}$ is the total cross section of the grain component $i$, $\epsilon_{\mathrm{H}_2,i}$ is the formation efficiency \citep{cazaux2004} on surfaces of the $i$-th component, and $S_\mathrm{H}$ is the sticking coefficient of the H atoms 
\begin{equation}\label{sticking}
S_\mathrm{H}=\left(1+0.04\left(\frac{T_g+T_d}{100}\right)^{1/2}+0.2\frac{T_g}{100}+0.08\left(\frac{T_g}{100}\right)^2\right)^{-1}   ,
\end{equation}
which depends on the gas and dust temperature, $T_g$ and $T_d$  \citep{hollenbach79}. The thermal velocity is $1.45\times 10^5(T_g/100)^{1/2}$~cm s$^{-1}$. 
The $i$-th dust cross section per H nucleus is given by
\begin{equation}\label{sigmadust}
\sigma_{d,i}=\int_{a_{i,-}}^{a_{i,+}}n_i(a)\pi a^2 da\, .
\end{equation}
Table~\ref{table:1} lists $\sigma_{d,i}$ for the different dust models.
\citet{cazaux2002} expressed the general expression for the $i$-th \HH\ formation efficiency:
\begin{figure}
\resizebox{\hsize}{!}{\includegraphics{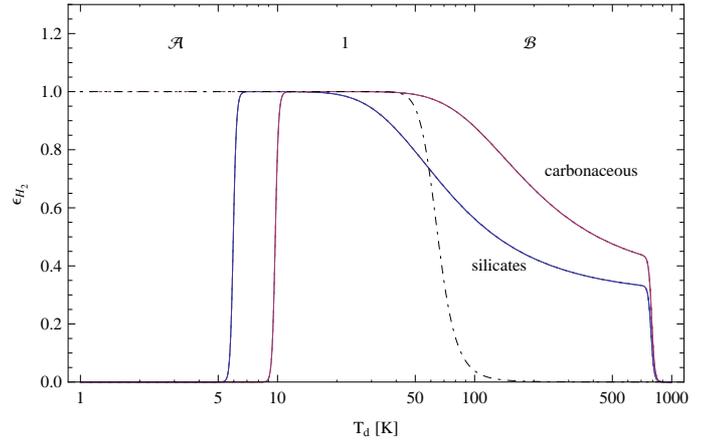}}
\caption{\HH\ formation efficiency for carbon and silicate surfaces. $\mathpzc{A}$ and $\mathpzc{B}$ denote the temperature regimes of the corresponding terms in Eq. (\ref{h2form}). The dash-dotted line denotes the standard formation efficiency given by \citet{hollenbach79}. }
\label{dustefffig}
\end{figure}

\begin{equation}
\epsilon_{\mathrm{H}_2,i}=\frac{\xi_i}{\mathpzc{A}+1+\mathpzc{B}}\, .\label{h2form}
\end{equation}
$\epsilon_{\mathrm{H}_2,i}$ is limited by three terms: the first term $\mathpzc{A}$  prohibits the evaporation of the newly formed molecules at low temperatures. The second term, equal to unity, dominates at higher grain temperatures; all incoming H atoms leave the surface as \HH. The third term $\mathpzc{B}$ governs the high-temperature regime, where evaporation of physisorbed atoms 
removes \ce{H} from the surface before it can recombine to \HH . It competes with the hopping of atoms into chemisorbed binding sites where the atoms can remain on the surface to form \HH\ at significantly higher temperatures.
The numerator $\xi_i$, describing the chemisorption, terminates the \HH\ formation at much higher temperatures, above several hundred Kelvin,
when even chemisorbed atoms start to evaporate.  

Figure \ref{dustefffig} shows $\epsilon_{\mathrm{H}_2}$ as a function of $T_d$ for silicate (blue line) and graphite (red line) surfaces.  
The old formation efficiency from \citet{hollenbach79} is shown as a dash-dotted line for comparison.
When applying this formation in the framework of the PDR model, see Fig. \ref{dustefffig}, the term
 $\mathpzc{A}$
turned out to be  
problematic because it shuts down the \HH\ release into the gas phase once $T< 10$K, accumulating all molecules on the grain surfaces. We question this term because Eq.~(\ref{h2form}) does not account for the \HH\ binding energy of 4.48 eV that upon formation is available to allow release into the gas phase even at very low temperatures. As a remedy we set
 $\mathpzc{A}=0$ to ensure efficient \HH\ formation even on cold dust surfaces (Cazaux (2011), priv. comm.)\footnote{All model results accounting for CT \HH\ formation shown in this paper assume $\mathpzc{A}=0$.}. More details on how to calculate the \HH\ formation efficiency are given in Appendix ~\ref{appendix:h2formation}. The total \HH\ formation rate can then be written as
\begin{equation}
\label{h2formtotal}
R_{\mathrm{H}_2}=\sum_{i=1}^\mathrm{NDS}R_{\mathrm{H}_2,i}=\frac{1}{2}n_\mathrm{H} v_H \sum_{i=1}^\mathrm{NDS} S_{\mathrm{H},i} n_{d,i} \sigma_{d,i}\epsilon_{\mathrm{H}_2,i}\, ,
\end{equation}

\begin{figure}
\resizebox{\hsize}{!}{\includegraphics{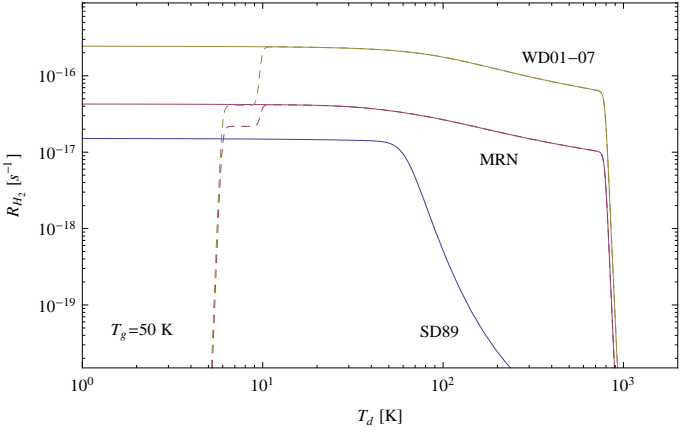}}
\caption{\HH\  formation rate $R_{\mathrm{H}_2}$ as a function of dust temperature $T_d$ for different formation models and dust compositions. The gas temperature $T_g$ is set to 50~K.  The dashed lines indicate the-low temperature formation in the original CT formation rates with $\mathpzc{A}\ne0$. SD89 is the formation rate from  \citet{SD89}.}
\label{Rdust1}
\end{figure}

\begin{figure}
\resizebox{\hsize}{!}{\includegraphics{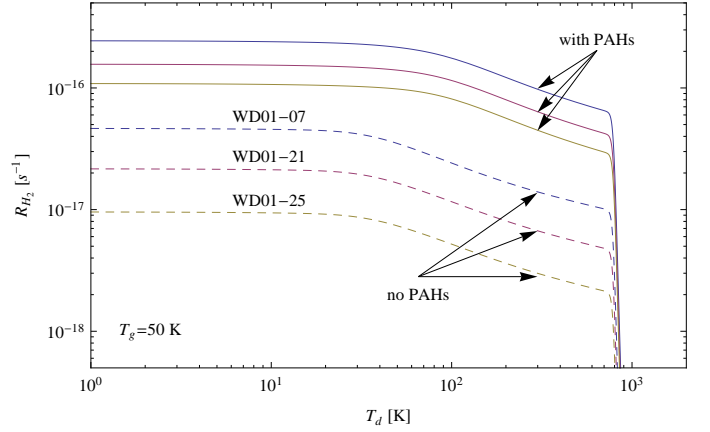}}
\caption{\HH\  formation rate $R_{\mathrm{H}_2}$ as a function of dust temperature $T_d$ for different dust compositions. The gas temperature $T_g$ is set to 50~K.  The dashed lines show the formation rate with suppressed \HH\ formation on PAHs. The solid lines show the formation rate allowing for \HH\ formation on PAH surfaces. }
\label{Rdust2}
\end{figure}

Fig.~\ref{Rdust1} compares the new and old \HH\ formation rates. The formation efficiency given by SD89 follows the treatment of \citet{hollenbach79}, which adopted a critical dust temperature of $T_{cr}\approx 65$~K above which the \HH\ formation efficiency drops below 0.5. \HH\ formation becomes inefficient above $\approx 100$~K. The \HH\ formation in the CT dust model remains efficient up to 1000~K.
The model by Cazaux\&Tielens allows a much more efficient \HH\ formation at higher temperatures. Additionally, the significantly larger dust surface in the MRN and WD01-7 model lead to a much higher \HH\ formation rate than SD89.

While accounting for chemisorbed H atoms significantly increases the \HH\ formation efficiency at higher surface temperatures, it is the total available grain cross section that distinguishes between different dust models. Table 
\ref{table:1} also compares the total dust cross section per H nucleus for the different dust compositions. The PAH cross sections contribute the majority to the final \HH\ formation rate. The resulting \HH\ formation rate for one H atom cm$^{-3}$ is shown in Fig. \ref{Rdust2}. Solid lines show the formation rate as a function of $T_d$ for a constant $T_g=50$~K for the WD01-7, WD01-21, and WD01-25 models allowing for \HH\ formation on the surface of PAHs. The dashed lines show the formation rates if the PAH surfaces are excluded from the \HH\ formation calculation. Consequently, the dust model with the largest effective cross section, WD01-7 shows the highest formation rate in both cases.  

\begin{figure}
\resizebox{\hsize}{!}{\includegraphics{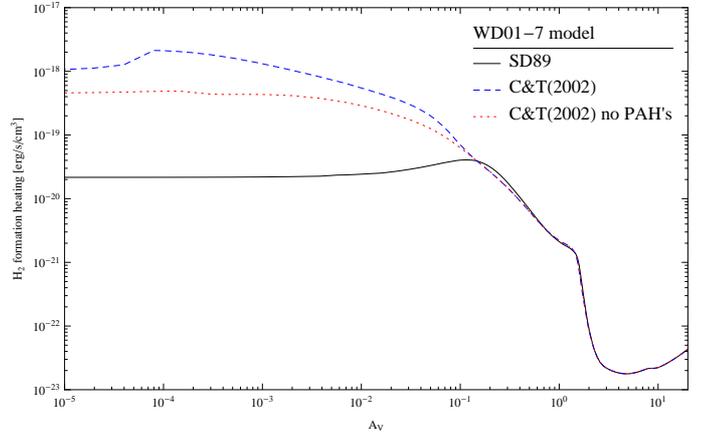}}
\caption{\HH\ formation heating rate  for a model clump of $M=$10 \msol, $n=10^5$~cm$^{-3}$, and $\chi=1000$.
 The different lines show the effect of different \HH\ formation treatment. All models have been computed using the WD01-7 dust properties regarding the radiative transfer and the PE heating. }
\label{h2formboth}
\end{figure}

The strong increase in the total \HH\ formation rate has an important effect in addition to the easier \HH\ production. Each formed \HH\ molecule releases its binding energy of 4.48 eV. A common assumption \citep[e.g.,][]{habart2004} is that this energy is equally used to a) release the molecule into the gas phase, b) increase the kinetic energy of the molecule, and for c) internal rotational-vibrational excitation. If we assume a uniform distribution, $1.5$ eV are available to heat the gas. 
Here we can measure the \HH\ formation rate across the cloud by equivalently looking at the \HH\ formation heating.

By including \HH\ formation according to Eq. (\ref{h2formtotal}) we also increase the  formation rate significantly at low values of $A_V$. This is shown in  Fig.~\ref{h2formboth}. All models in the plot were computed using the WD01-7 dust properties regarding the radiative transfer and the PE heating. The solid line shows the results for the formation rate from \citet{SD89}. The result for the full \HH\ formation treatment from CT (with $\mathpzc{A}=0$), suppressing and allowing \HH\ formation on PAH surfaces, are shown by the dotted and dashed line. The much larger surface area of the WD01-7 dust model leads to a significant increase of the formation rate. Particularly the large PAH surface in that dust model contributes dominantly to the total grain surface and consequently to the total \HH\ formation heating.

It is important to note that the \HH\ formation heating is extensive at cloud depths where most of the hydrogen is in the form of \ce{H}. This leads to the curious situation of a strong \HH\ formation heating in the absence of \HH\ molecules. The gas temperatures at these depths can reach more than thousand Kelvin. Under these conditions two more \HH\ destruction reaction become important: 
\reaction[h2disskin1]{H2 + H -> H + H + H }
\reaction[h2disskin2]{H2 + H2 -> H2 + H + H . }
Each destruction process requires 4.48 eV to break the binding, in contrast to the release of 4.48 eV during the formation. The much more efficient formation of molecular hydrogen by Cazaux \& Tielens makes it necessary to adopt an additional cooling term in the chemical network, accounting for the 4.48 eV energy consumption during the kinetic dissociation  \citep{lepp83}. The cooling rate can be written as
\begin{equation}
\Lambda_{kd}=7.2\times 10^{-12}n_{\mathrm{H}_2}\left(k_{(\mathrm{\ref{h2disskin1}})}n_\mathrm{H}+k_{(\mathrm{\ref{h2disskin2}})}n_{\mathrm{H}_2}\right)\,\mathrm{erg\, s}^{-1}\mathrm{cm}^{-3}\, ,
\end{equation}
where $k_{(\mathrm{\ref{h2disskin1}})}$ and $k_{(\mathrm{\ref{h2disskin2}})}$ are the temperature-dependent rate coefficients for reactions (\ref{h2disskin1}) and (\ref{h2disskin2}).

The role of interstellar dust in ISM physics and chemistry remains one of the key problems in modern astrophysics. In this section we demonstrate the strong effect of \HH\ formation on PAH surfaces on the chemistry and the thermal conditions in PDRs. However, it remains unknown whether PAHs really contribute to the formation of molecular hydrogen. Naturally, this is connected to the uncertainty about the detailed form and distribution of interstellar PAHs.

By assuming that \ce{H2} formation can take place on PAH surfaces the formation rate from Cazaux \& Tielens predicts rates of a few times $10^{-16}$~s$^{-1}$ when applied to the dust distributions presented by WD01. This is about a factor 10 higher than what has commonly been found for the diffuse medium \citep{jura1974,browning2003,gry2002,welty2003,wolfire2008}. However, the usual values of $R \backsim 2-4 \times 10^{-17}$~cm$^{3}$~s$^{-1}$ are mean values along the line of sight while our results are local values and a direct comparison is difficult. Recently, \citet{lebourlot2012} presented an update to the \ce{H2} formation formalism in the Meudon PDR code and also found a strong increase of the local \ce{H2} formation rates. If we prevent \ce{H2} formation on PAH surfaces,  the total grain surface available to  form molecular hydrogen is significantly reduced and we predict formation rates of the same order as found in the diffuse ISM. However, PDR models assuming diffuse gas formation rates are not able to reproduce the observed extent of \ce{H2} line excitation \citep{habart2011}. \ce{H2} formation on PAHs has also been proposed by \citet{habart2003} to explain the \ce{H2} emission in $\rho$~Oph, and recently \citet{mennella2012} concluded from experimental studies that hydrogenated neutral polycyclic aromatic hydrocarbon molecules act as catalysts for the formation of molecular hydrogen.

\subsection{Photoelectric heating\label{sect:PEH}}

 \citet{WD01PEH} presented the photoelectric heating rate for the WD01 grain size distributions in a parametrized form as a function of $\psi=(G\sqrt{T})/n_e$, where $T$ is the gas temperature in Kelvin and $\psi$ is in units of K$^{1/2}$cm$^3$. To convert the FUV field from Habing $G$ to Draine $\chi$ units use $G=1.71 \chi$. These parametrizations are fairly accurate when $10^3\mathrm{K}\le T\le 10^4 K$ and $10^2\mathrm{K}^{1/2}\mathrm{cm}^3\le \psi\le 10^6 \mathrm{K}^{1/2}\mathrm{cm}^3$. However, the parameter range in PDR models allows  $\psi$ to  reach values as high as $10^9 \mathrm{K}^{1/2}\mathrm{cm}^3$ and as low as $10^{-2} \mathrm{K}^{1/2}\mathrm{cm}^3$. \citet{weingartner09} provided updated calculations of $\Gamma_\mathrm{pc}$ and $\Lambda_\mathrm{gr}$ for the dust distributions WD01-7, WD01-21, and WD01-25 up to  $\psi=10^9 \mathrm{K}^{1/2}\mathrm{cm}^3$. These rates are fairly well reproduced by the new analytic fits:
\begin{equation}
\Gamma_\mathrm{pc}= G n_\mathrm{H}\left(\frac{C_0+C_1 T^{C_4}}{1+C_2\psi^{C_5}[1+C_3\psi^{C_6}]}+C_7\right)\,10^{-26}\,\mathrm{erg}\,\mathrm{s}^{-1}\,\mathrm{cm}^{-3}\label{PEHnew}
\end{equation}
and
\begin{eqnarray}\label{lamdanew}
\Lambda_\mathrm{gr}&=&n_e n_\mathrm{H} \left(D_1 \left(D_2 T^{D_3}+D_4\right) T^{D_0}\right .\\ \nonumber
&& \left(D_5 e^{D_6 (x-10.3) \left(-T^{D_7}\right)}+1\right)^{-D_5}\\ \nonumber
&&\left .+T^{D_8}+D_9 T^{D_{10}}+D_{11}\right)\\ \nonumber
&& 10^{-28}\,\mathrm{erg}\,\mathrm{s}^{-1}\,\mathrm{cm}^{-3}.
\end{eqnarray}
Numerical values for $C_i$ and $D_i$ are tabulated in the appendix in Table \ref{tab1} and \ref{tab2}.

\begin{figure}
\resizebox{\hsize}{!}{\includegraphics{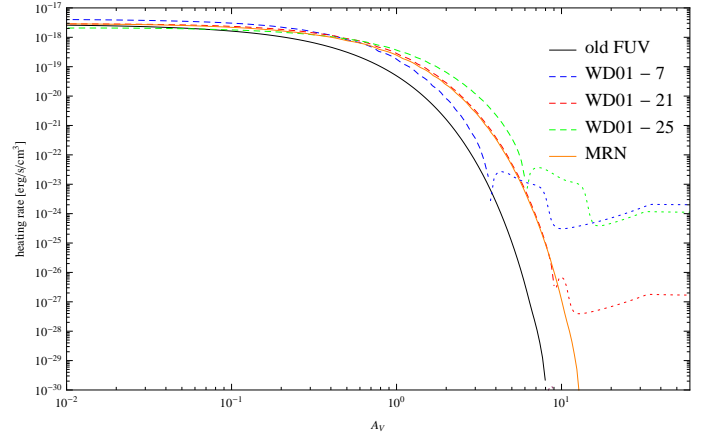}}
\caption{Photoelectric heating rate for a model clump of $M=$10 \msol, $n=10^5$~cm$^{-3}$, and $\chi=1000$. The different lines denote different dust compositions. Models with an MRN dust composition apply the PE heating efficiency as given by \citet{bt94}. Dotted lines indicate a shift from heating to cooling because of $\Lambda_\mathrm{gr}>\Gamma_\mathrm{pc}$.}
\label{PEH}
\end{figure}

Figure \ref{PEH} shows the PE heating rates for a model clump of $M=$10 \msol, $n=10^5$~cm$^{-3}$, and $\chi=1000$. The solid lines show the photoelectric heating (PEH) rates using the method described by \citet{bt94}. The PEH rate obtained for an MRN dust composition 
when solving the full FUV radiative transfer problem
is enhanced compared to the old \ktau\ values because of the stronger FUV intensities due to the scattering.  The dashed lines in Fig. \ref{PEH} show the heating rates using the updated parametrizations from Eqs. (\ref{PEHnew}) and (\ref{lamdanew}) for the different dust models.  Dotted lines indicate negative values, i.e., a transition from heating to cooling because $\Lambda_\mathrm{gr}>\Gamma_\mathrm{pc}$. For $A_V>1$ the PEH rate of the WD01 dust models follow the FUV intensity of the respective models, i.e., the model with the highest mean FUV intensity, WD01-25, shows the strongest PEH rate while the model WD01-07 drops quickest. However, it is interesting to note that for $A_V<0.1$ the PEH rate for the WD01-07 dust produces the strongest PEH rate, leading to the highest gas temperature of all models (see Fig. \ref{Tgas}).
    
\section{Application of the single clump model\label{sect:application}}
\subsection{Impact on gas temperature\label{sect:gastemp}}
\begin{figure*}
\includegraphics[width=17cm]{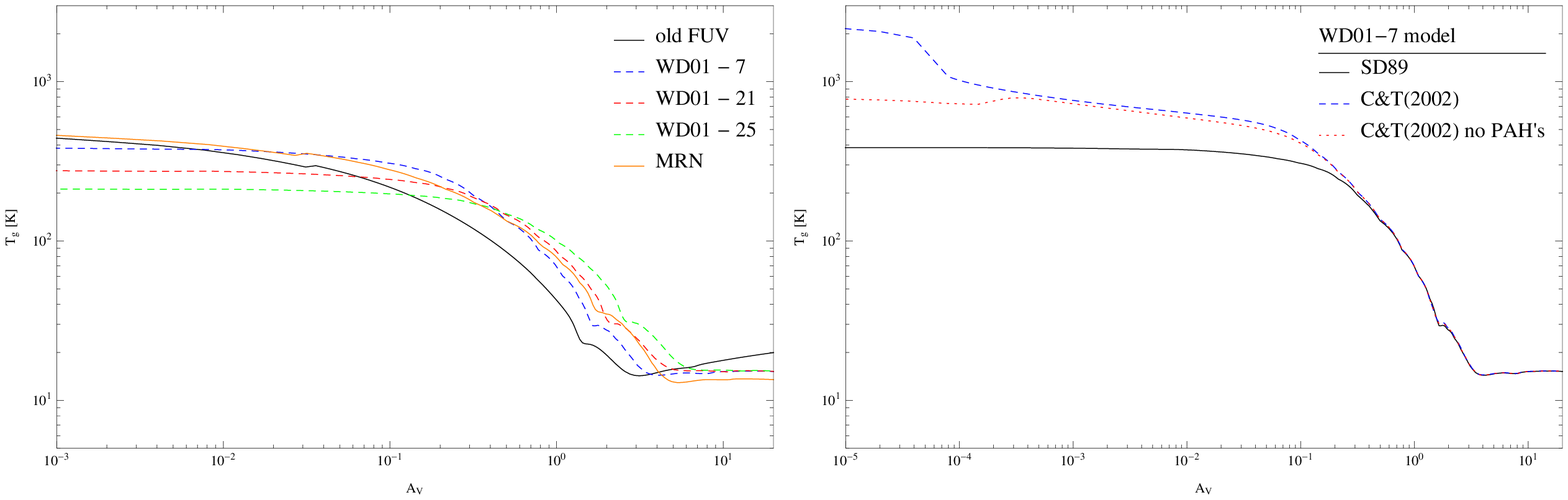}
\caption{Gas temperature $T_g$ for a model clump of $M=$10 \msol, $n=10^5$~cm$^{-3}$, and $\chi=1000$. {\bf Left panel:} The different lines denote different dust models that affect the FUV radiative transfer and the PE heating. All models in the left panel werecalculated using the SD89 \HH\ formation. {\bf Right panel:} The different lines show the effect of different \HH\ formation treatment. All models in the right panel were computed using the WD01-7 dust properties regarding the radiative transfer and the PE heating. }
\label{Tgas}
\end{figure*}

The changes in computating the PE heating and the \HH\ formation, which depend on the applied dust model, have a profound effect on the thermal structure of the model clump. Figure \ref{Tgas} shows the separate effects of the changed FUV and resulting PE heating (left panel) and  different \HH\ formation treatment (right panel). Among the different dust models presented by \citet{WD01} the WD01-7 dust models shows the highest total PE heating rate because it contains the largest population of very small grains \citep{WD01PEH}. Depending on  $\psi$, the PE heating rate from \citet{bt94} can become higher than in the WD01 models despite its smaller population of very small grains. This is because they assume a higher electron sticking coefficient. In the chemically very active zone of $0.1<A_V<2$, the new FUV treatment leads to systematically warmer gas temperatures. The net effect of the dust model on the PE heating strongly depends on the applied dust model, with the PE heating efficiency  proportional to the size of the very small grain population.
A second effect comes from the \HH\ formation heating. The right panel in Fig. \ref{Tgas} shows how the increase in the total available grain surface raises the gas temperature. However, this mostly affects the outer layers of a model clump where the high grain temperatures require accounting for chemisorbed H atoms to allow the formation of molecular hydrogen.

Overall, we can distinguish two regions: For optical depths $A_V \ga 0.5$, the different attenuation of the FUV field by different dust distributions dominates the gas temperature. The model with the lowest reddening, i.e., the highest FUV intensity at those depths, shows the highest gas temperature. At lower optical depths, the FUV intensity is in all cases high enough that the total dust surface available for PE heating and H$_2$ formation determines the gas temperature. Here, the models with PAHs and many small grains, which also produce a higher reddening, show the highest gas temperature.
 
\subsection{Impact on gas chemistry\label{sect:gaschem}}

The described changes to the model have a significant effect on the chemical structure of the cloud. The influence of the assumed dust composition and dust size distribution affects the FUV radiative transfer and the improved treatment of dust scattering increases the FUV intensity throughout the clump compared to the original approximation. Chemical species that are dominantly formed or destroyed via photo-ionization or photo-dissociation processes are affected. Secondly, the stronger \HH\ formation efficiency leads to an increased \HH\ formation heating contribution, which in turn produces an increase in gas temperature at low $A_V$. 
Consequently, the abundance of the species that are dominantly formed in the outer regions of the molecular clump is changed because the higher gas temperature accelerates their respective formation and destruction reactions in the gas phase.
\begin{figure}
\resizebox{\hsize}{!}{\includegraphics{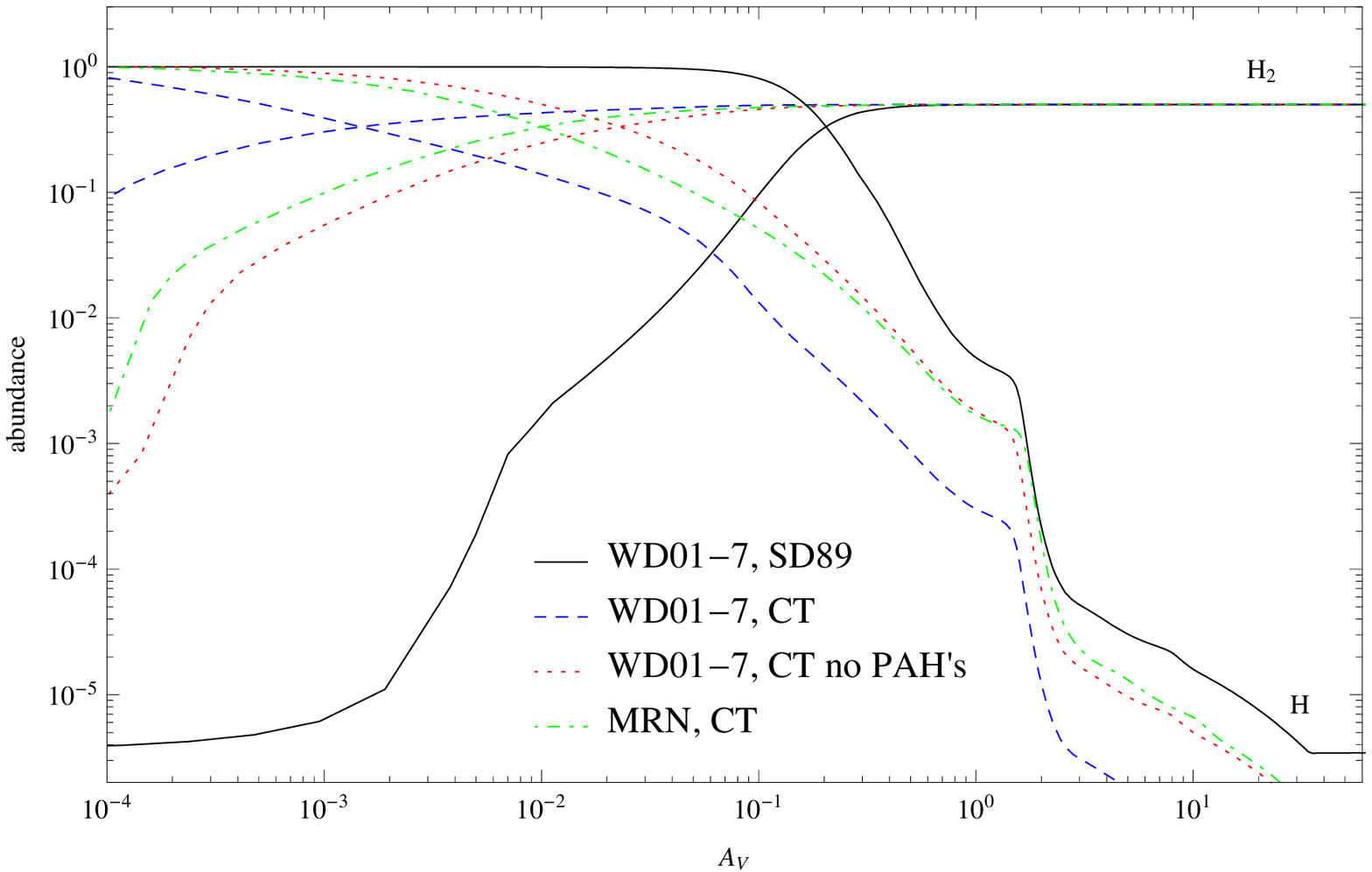}}
\caption{H and \HH\ abundances for a model clump of $M=$10 \msol, $n=10^5$~cm$^{-3}$, and $\chi=1000$. The different lines denote different treatments of the \HH\ formation. The solid line shows a model using the standard \HH\ formation rate as given by \citet{SD89}. The dashed and dotted lines show model results for \HH\ formation according to CT where the formation on PAH surfaces is switched on and off, respectively. All three models assume a WD01-7 dust distribution. The green dash-dotted line assumes an MRN dust model and uses the CT \HH\ formation. }
\label{chemHH2form}
\end{figure}
\begin{figure*}
\centering
\includegraphics[width=17cm]{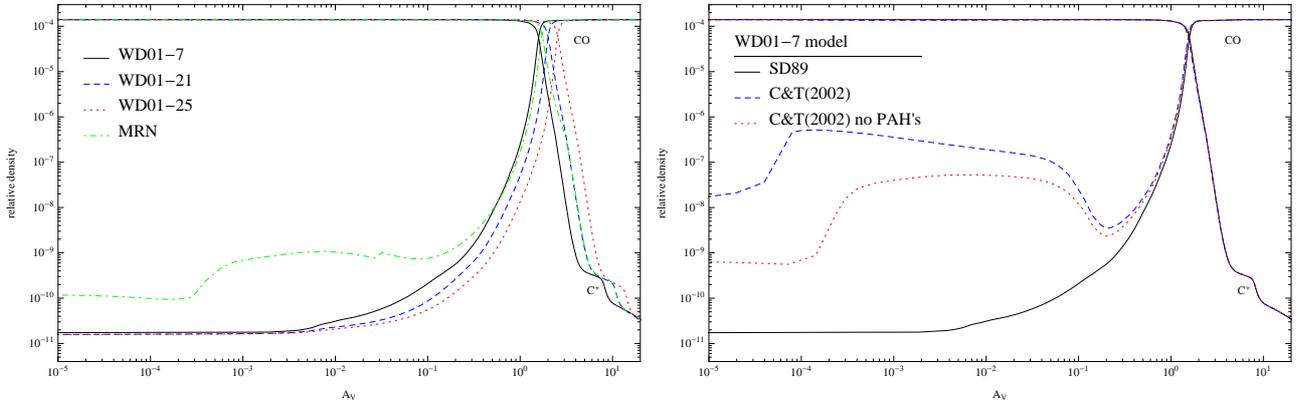}
\caption{CO and \Cp\  abundances for a model clump of $M=$10 \msol, $n=10^5$~cm$^{-3}$, and $\chi=1000$. {\bf Left panel:} The different lines denote different dust models that affect the FUV radiative transfer and the PE heating. All models in the left panel have been calculated using the SD89 \HH\ formation. {\bf Right panel:} The different lines show the effect of different \HH\ formation treatment. All models in the right panel have been computed using the WD01-7 dust properties regarding the radiative transfer and the PE heating. }
\label{chemCpCO}
\end{figure*}

Both effects can go in opposite directions, but usually one effect dominates. In Fig. (\ref{chemHH2form}) we show the dependence of the H-\HH\ transition zone on the applied \HH\ formation rate. We compare three models assuming a WD01-7 dust model with full radiative transfer and photoelectric heating treatment. The only difference between the model calculations is the treatment of \HH\ formation. We apply (1) the standard \HH\ formation rate as given by \citet{SD89} (black, solid line), (2) the \HH\ formation given by CT with \HH\ formation taking place on the PAH surface (blue, dashed line), (3) and the \HH\ formation given by CT with \HH\ formation on PAHs suppressed (red, dotted line). \HH\ chemistry is unique in the sense that it only weakly depends on the gas temperature (see Eqs. (\ref{h2rate}, \ref{sticking})). Figure \ref{chemHH2form} shows how the different treatment of the \HH\ formation influences the details of the H-\HH\ transition.
In that particular model clump, the dust temperature is about 50 K and almost identical in all models, leading to a \HH\ formation efficiency of unity across the dust models shown. The only significant difference is the total dust surface of the various models.
The SD89 approximation, case (1), has the smallest available surface ($\sigma_d=4.14\times 10^{-22}$~cm$^2$)
 followed by CT without PAHs and CT including PAHs (see Table \ref{table:1}). This  agrees with the H-\HH\ transition moving outwards with growing dust surface in Fig. \ref{chemHH2form}. 
To emphasize this behavior, we added a fourth dust model to the plot, showing the results for the MRN model. The total surface in the MRN model is only slightly smaller than the WD01-7 model without PAHs and consequently, both models show a very similar behavior.  Please note that the contribution of graphite and silicate to the total dust surface is still very different in the different dust models. The MRN model possesses about equal surfaces in both kinds, while the majority of the big grain surfaces in the WD01-7 model comes from silicates.

In contrast to \HH\ , all other chemical species in the gas phase are affected by variations of the local FUV intensity and the local gas temperature.  This is apparent, for example,  in the \Cp\ to CO transition. 
Figure \ref{chemCpCO} shows the density profile of C$^+$ and CO for different model calculations. In the left panel we show the influence of the different dust models in terms of radiative transfer and PE heating, while the right panel demonstrates how the different \HH\ formation treatment influences the chemistry.
 All models in the left panel were calculated using the SD89 \HH\ formation.
  The transition from \Cp\ to CO occurs at different $A_V$ for all models. Consistent with the FUV intensities shown in Figs. \ref{FUV2depth1} and \ref{FUV2depth2}, the models with the lowest FUV intensities show a C$^+$ to CO transition at lowest $A_V$, while models with a weak dust attenuation, e.g. WD01-25, exhibit the same transition deeper in the cloud. Additionally, we note an enhanced CO density at low $A_V$ in the WD01-7 model. This is a result of the higher gas temperature due to the stronger PE heating efficiencies of that dust model. 
\begin{figure}
\resizebox{\hsize}{!}{\includegraphics{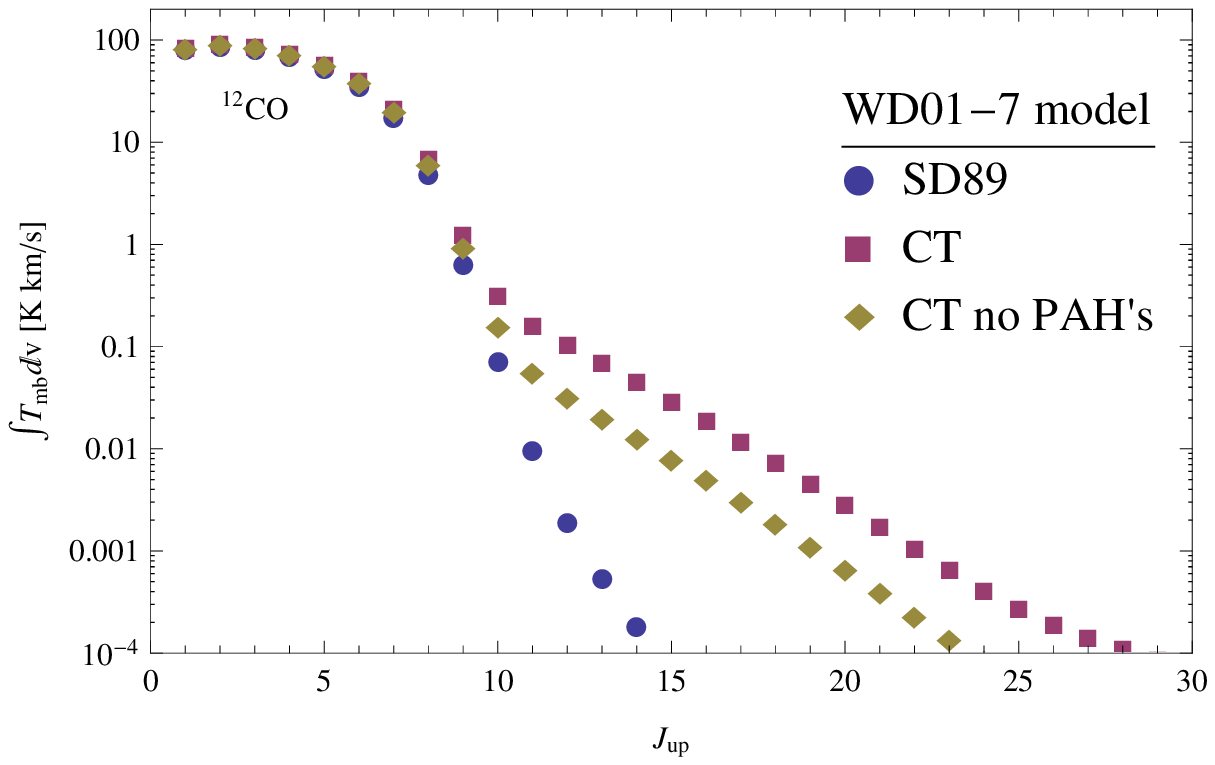}}
\caption{Total line integrated clump-averaged intensities of $^{12}$CO transitions for a model clump of $M=10$  \msol, $n=10^5$~cm$^{-3}$, and $\chi=1000$. The different lines show the effect of different \HH\ formation treatment. All models were computed using the WD01-7 dust properties regarding the radiative transfer and the PE heating. }
\label{chemCOladder}
\end{figure}

The right panel in Fig. \ref{chemCpCO} shows the isolated effect of the different \HH\ formation treatment. All models in the right panel were computed using the WD01-7 dust properties regarding the radiative transfer and the PE heating. The \Cp-CO transition is similar across all shown models, since the FUV intensity, which controlls the photo-dissociation of CO, is the same across all models. However, the CO abundance at lower $A_V$ shows very large differences. The models using the CT dust formation scheme produce more CO at the cloud edge compared to the SD89 \HH\ formation. The CT models provide a stronger \HH\ formation heating contribution. The additional heating term produces a significant population of hot CO in the outer layers of the model clump.

The production of hot, i.e., strongly excited, CO is visible in the spectral line emission of the model clumps. In Fig. \ref{chemCOladder} we show the corresponding total CO line emission of the model clumps shown in the right panel in Fig. \ref{chemCpCO}. For transitions higher than J=8-7, the line-integrated intensities start to show significant differences for the three models. In this particular case, the strength of the high-$J$ CO transitions is proportional to the total {\HH}-forming grain surface. The outer layers of the model clumps contribute significantly to the total CO line emission of this model. As a consequence, any modeling with a large surface of the clouds, e.g., assuming a clumpy structure, will show much stronger high-$J$ CO emission lines compared to non-clumpy model attempts.

\begin{figure*}
\centering
\includegraphics[width=17cm]{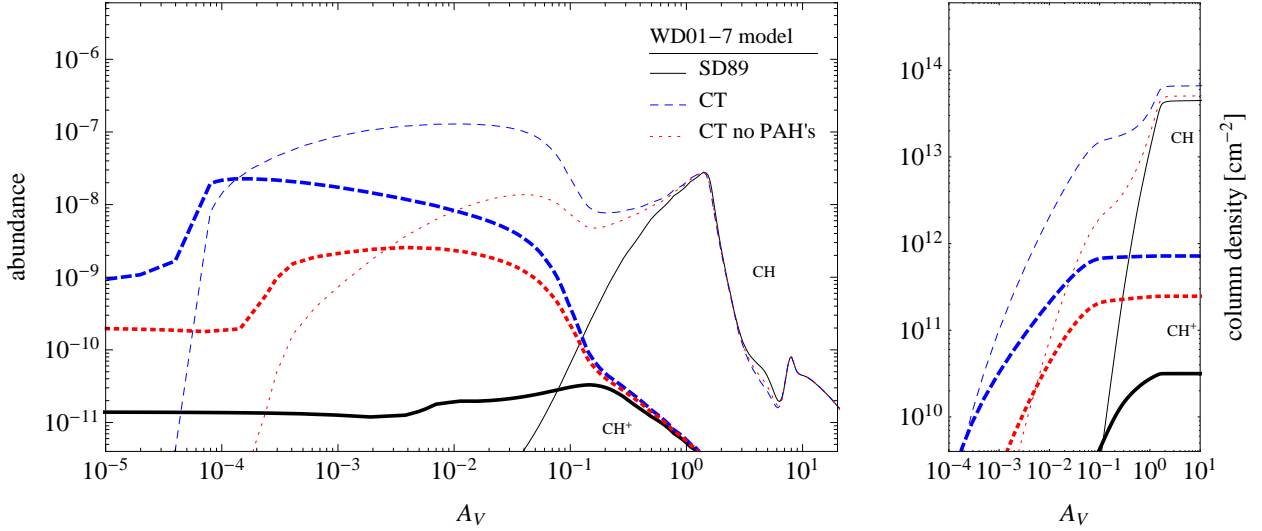}
\caption{\ce{CH} and \ce{CH+} abundances (left) and column densities (right)  for a model clump of $M=10$  \msol, $n=10^5$~cm$^{-3}$, and $\chi=1000$.The different lines show the effect of different \HH\ formation treatment. All models were computed using the WD01-7 dust properties regarding the radiative transfer and the PE heating. }
\label{chemCH}
\end{figure*}
\begin{figure*}
\centering
\includegraphics[width=17cm]{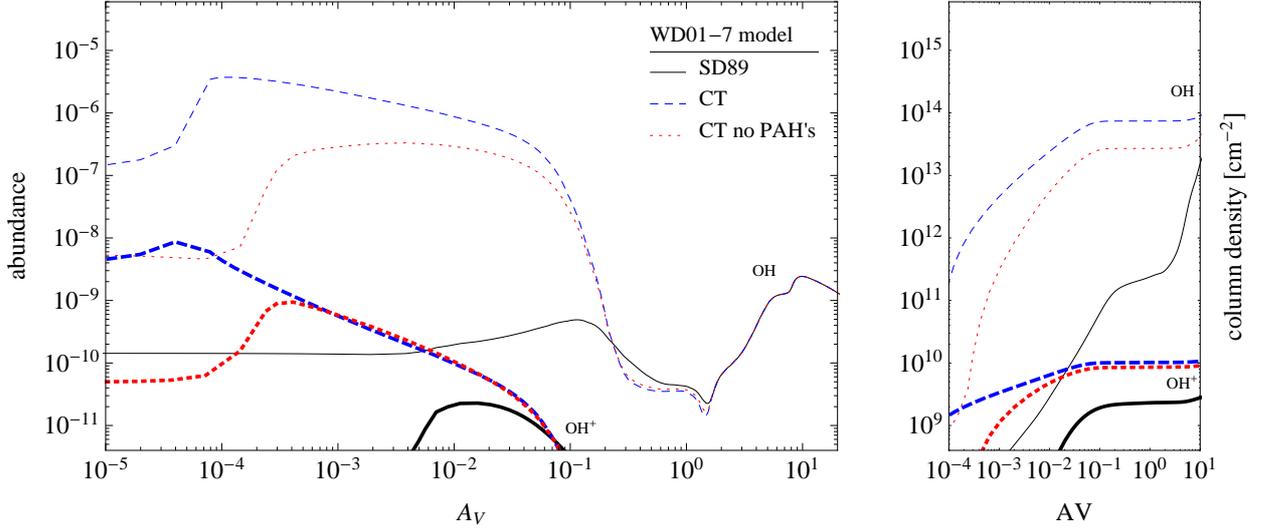}
\caption{\ce{OH} and \ce{OH+} abundances (left) and column densities (right)  for a model clump of $M=10$  \msol, $n=10^5$~cm$^{-3}$, and $\chi=1000$.The different lines show the effect of different \HH\ formation treatment. All models were computed using the WD01-7 dust properties regarding the radiative transfer and the PE heating. }
\label{chemOH}
\end{figure*}

As an additional example of chemical species that predominantly form in the outer regions of molecular clouds and thus are affected by the \HH\ formation treatment we show in Fig. \ref{chemCH} the \ce{CH} and \ce{CH+} abundances and column densities and in Fig. \ref{chemOH} the \ce{OH} and \ce{OH+} abundances and column densities. All these species are significantly affected by the strong \HH\ formation heating effect in the outer parts of the model clump. The column density effect is weakest for the CH because of the abundance peak at $A_V\approx 1$. The other three species show a stronger effect on the column density because of the weak formation in deeper clump regions. Recent Herschel observations show high column densities of these surface tracers \citep[e.g.][]{qin2010, falgarone2010}, which were difficult to reproduce in existing PDR models. A more efficient \HH\ formation modeling with the accompanying formation heating effect may help to resolve the discrepancy between models and observations.

\subsubsection{Full model SED\label{sect:SED}}

 \begin{figure}
 \resizebox{\hsize}{!}{\includegraphics{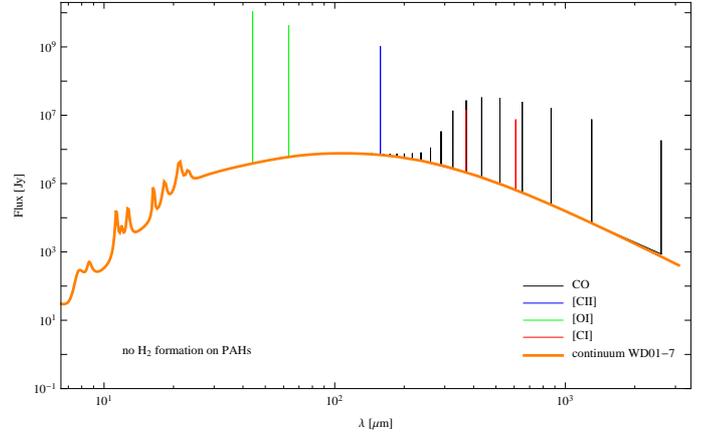}}
 \resizebox{\hsize}{!}{\includegraphics{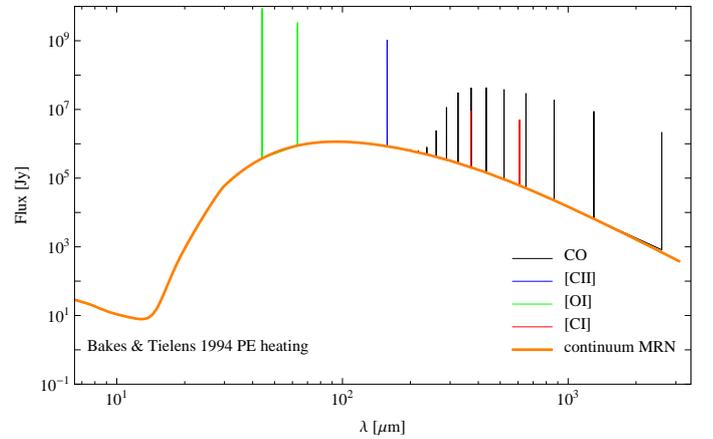}}
 \caption{Full continuum and spectral line flux emitted by a model clump of $M=10$  \msol, $n=10^5$~cm$^{-3}$, and $\chi=1000$. The two panels show the results for different dust models.
 {\bf Top:} WD01-7 dust model, equivalent to a $R_V=3.1$. This is the dust model with the largest grain surface contributing to the heating.
{\bf Bottom:} MRN dust mode. This is a dust model with a slightly smaller grain surface compared to WD01-7. The PE heating was calculated according to \citet{bt94}. }
 \label{fullSLED}
 \end{figure}

\begin{table*}[htb]
\begin{center}
\caption{Line emission $\int I dv$ for the three models shown in Fig. \ref{fullSLED}.}
\label{table:2}
\begin{tabular}    {ll r l l l l l }
\hline\hline
&line&$\lambda$ &WD01-7&WD01-7&WD01-25&WD01-25&MRN\\
&&&w/o PAHs&with PAHS&w/o PAHS&with PAHS&\\
&&[$\mu$m]&[erg/s/cm$^2/sr$]&[erg/s/cm$^2/sr$]&[erg/s/cm$^2/sr$]&[erg/s/cm$^2/sr$]&[erg/s/cm$^2/sr$]\\ \hline
$I_\mathrm{TIR}$&&&0.290&0.290&0.286&0.280&0.278\\
\CII\ & $^2P_{3/2}- {^2P_{1/2}}$&158&$7.5\times 10^{-4}$&$7.5\times 10^{-4}$&$9.2\times 10^{-4}$&$9.3\times 10^{-4}$&$7.7\times 10^{-4}$\\
\OI& $^3P_1- {^3P_2}$ &63&$6.8\times 10^{-3}$&$6.6\times 10^{-3}$&$5.7\times 10^{-3}$&$6.0\times 10^{-3}$&$5.5\times 10^{-3}$\\
\OI& $^3P_0- {^3P_1}$&146&$5.0\times 10^{-4}$&$5.1\times 10^{-4}$&$4.2\times 10^{-4}$&$4.9\times  10^{-4}$&$4.1\times 10^{-4}$\\
\CI& $^3P_1- {^3P_0}$&609&$1.2\times 10^{-6}$&$1.2\times 10^{-6}$&$1.4\times 10^{-6}$&$1.3\times 10^{-6}$&$8.2\times 10^{-7}$\\
\CI& $^3P_2- {^3P_1}$&370&$3.9\times 10^{-6}$&$3.9\times  10^{-6}$&$4.6\times 10^{-6}$&$4.2\times 10^{-6}$&$2.4\times 10^{-6}$\\
CO& (1--0)&2601&$1.3\times 10^{-7}$&$1.3\times 10^{-7}$&$1.2\times 10^{-7}$&$1.3\times 10^{-7}$&$1.5\times 10^{-7}$\\
CO& (4--3)&650&$7.3\times 10^{-6}$&$7.4\times 10^{-6}$&$7.0\times 10^{-6}$&$7.5\times 10^{-6}$&$8.9\times 10^{-6}$\\
CO& (7--6)&372&$1.1\times 10^{-5}$&$1.1\times 10^{-5}$&$1.3\times 10^{-5}$&$1.5\times 10^{-5}$&$1.9\times 10^{-5}$\\
CO& (10--9)&260&$2.7\times 10^{-7}$&$5.0\times   10^{-7}$&$2.8\times 10^{-7}$&$4.5\times 10^{-7}$&$7.5\times 10^{-7}$\\
CO& (15--14)&174&$5.6\times 10^{-8}$&$1.6\times 10^{-7}$&$2.4\times 10^{-10}$&$2.5\times 10^{-8}$&$8.4\times 10^{-9}$\\
CO& (20--19)&130&$1.2\times 10^{-8}$&$3.7\times 10^{-8}$&$8.7\times 10^{-12}$&$7.3\times 10^{-9}$&$1.7\times 10^{-9}$\\
\hline
\end{tabular}
\end{center}
\end{table*}
In Fig. \ref{fullSLED} we show the full continuum and spectral line flux (in units of Jy) emitted by a model clump of $M=10$  \msol, $n=10^5$~cm$^{-3}$, and $\chi=1000$. The two panels show the results for different dust models. The top panel shows the result for the dust model with the largest effective surface contributing to the gas heating, the WD01-7 model, equivalent to a $R_V=3.1$. The strong gas heating is visible in the large population of intense high-$J$ CO emission lines with J>10 stemming from the outer layers of the cloud (see Table~\ref{table:2}).  \footnote{The level population of the CO energy levels was computed up to J=50. }
In the bottom panel we show the result for a MRN dust model, visible through the lack of PAH emission features in the NIR. 

In Table \ref{table:2} we summarize the line emission of the WD01-7 and WD01-25 (with and without \HH\ formation on PAHs), as well as the MRN models quantitatively. 
The differences between the WD01-7 and WD01-25 models are limited to the line emission, the continuum emission is only marginally different.
The WD01-25 models show weaker high-$J$ CO emission because of the smaller grain surface and correspondingly a weaker \HH\ formation heating. 
The high-$J$ CO emission of the MRN model falls between the WD01-7 and the WD01-25 dust models without \HH\ formation on PAHs. 
 Allowing for \HH\ formation on PAHs will add a strong heating contribution from the formation processes on the large PAH surface, exciting the high-$J$ CO transitions by a large factor.

\section{Summary\label{sect:summary}}

We revised the treatment of the dust in the \ktau\ PDR model code to achieve a consistent description of the dust related physics in the code. The computation of dust properties in numerical models of photo-dissociation regions afflicts the chemical and physical structure of a model cloud via multiple effects. The major areas where dust properties play an important role are
\begin{enumerate}
\item{the optical dust properties, i.e., their influence on the radiative transfer}
\item{the dust temperature (neglecting non-equilibrium heating for the very small grains), which influences surface chemistry, most importantly the formation of molecular hydrogen}
\item{the heating and cooling capabilities of dust grains via photoelectric heating and gas-grain collisional cooling.}
\end{enumerate}
The effect of changes in these areas on the physical and chemical structure of a model can be profound and we described their respective influence in detail. 

We notice two opposite effects: Increasing the fraction of small grains steepens the slope of the extinction curve so that the FUV intensity drops faster when going into the cloud, which leads to a temperature decrease at high optical depths. However, it also provides more surface for PE heating and H$_2$ formation, thus increasing the heating efficiency. As a consequence, the gas becomes hotter in the outer layers, but colder inside the cloud when the fraction of small grains and PAHs is increased. Only tracers of the outer layers are able to reliably measure the gas heating efficiency. In contrast to common wisdom, this is not even true for ionized carbon [CII], which traces both regimes.

The most influential modification of the code is the treatment of \HH\ formation on the surface of dust grains. Allowing for chemisorption, i.e., including the Eley-Rideal effect into the formulation of the \HH formation efficiency, significantly increases the overall formation rate of molecular hydrogen. If a significant fraction of the dust consists of very small grains and PAHs, their very large contribution to the total grain surface leads to an additional enhancement of the total \HH\ formation. If the molecule formation is enhanced, the formation heating is enhanced as well, providing a profound heating contribution in the outer layers of the model clumps, which significantly affects the local chemistry.

\begin{acknowledgements} 
This work was supported by the German
      \emph{Deut\-sche For\-schungs\-ge\-mein\-schaft, DFG\/} project
      number Os~177/1--1. R.S. acknowledges  support partly from the Polish NCN grant 2011/01/B/ST9/02229.
We acknowledge the use of the UDfA (a.k.a. UMIST)
(http://www.udfa.net/) chemical reaction databases.
Some kinetic data we used were downloaded from the online 
database KIDA (KInetic Database for 
Astrochemistry, http://kida.obs.u-bordeaux1.fr).
We thank the anonymous referee for many remarks that significantly improved this paper. 
We would like to thank J.~Weingartner for the calculation of the photoelectric heating rates and collisional cooling rates for the extended parameter range and for the helpful remarks on the calculation of the total heating rate.  
\end{acknowledgements}

\bibliographystyle{aa}
\bibliography{ref}

\appendix
\section{Isotropic vs. non-isotropic scattering\label{scattering}}
\begin{figure*}[htb]
 \centering
 \includegraphics[width=17cm]{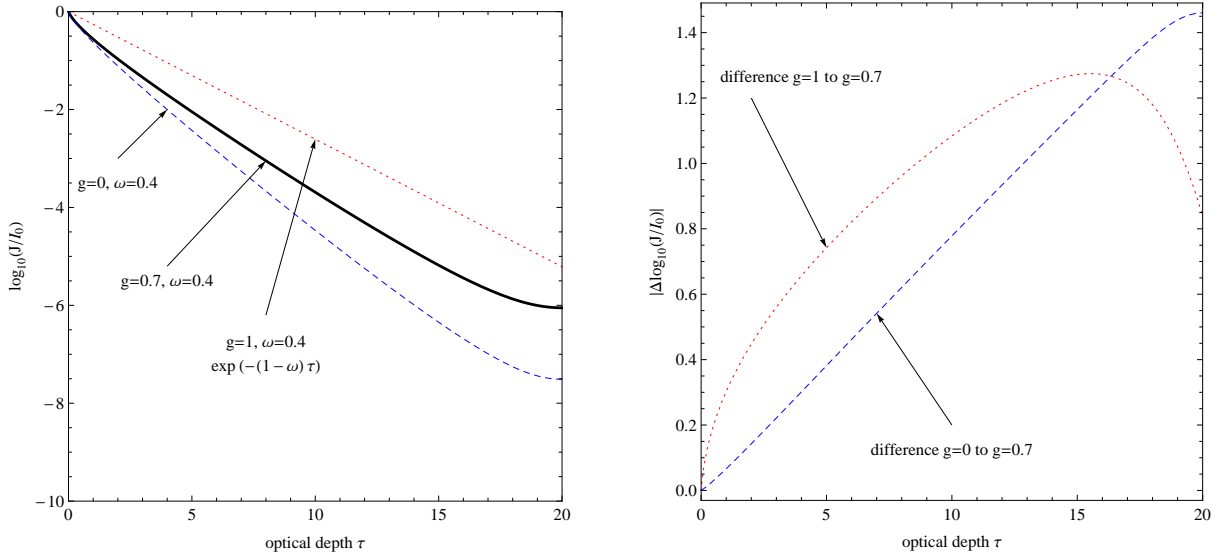}
\caption{
{\bf Left panel}: Mean intensity versus optical depth $\tau$ normalized to the incident specific intensity $I_0$. The solid black curve shows the depth dependence of the mean intensity assuming dust parameters $\omega=0.4$, $g=0.7$, approximately describing the WD01-7 dust sort. The dashed and dotted curves denote mean intensities using the same albedo $\omega$,  but assuming isotropic and pure forward-scattering, respectively. {\bf Right panel:} Absolute error $|\log(J_i/I_0)-\log(J_k/I_0)|$ of assuming isotropic $g=0$ and pure forward-scattering $g = 1$ with respect to more realistic values $\omega=0.4$, $g=0.7$ (dashed and dotted lines, respectively).} \label{scatter}
 \end{figure*}

The MCDRT code assumes isotropic scattering, i.e., a mean scattering angle $g=\langle \cos \Theta\rangle=0$ when calculating the frequency-dependent FUV radiative transfer.  However, taking into account the detailed material properties and size distribution of interstellar dust results in  different values of $g$. 
Reasonable values for interstellar dust in the FUV range fall around $g\approx 0.7$ and $\omega\approx 0.4$ \citep{li01b}. In the following we show that 
even in this case of strong forward-scattering the isotropic scattering treatment  poses a clear improvement compared to the pure forward-scattering cased used in our previous model.

To quantify the error
we used the {\it spherical harmonics} method, presented by \citet{FlanneryRoberge1980} to calculate the penetration of UV radiation of a spherical cloud. Applying their method, we  calculated the attenuation of the mean intensity as a function of optical depth for a given asymmetry factor $g$ and albedo $\omega$. For more details on the method  see also \citet{roberge1983,lepetit2006,goicoechea2007}.

In Fig.~\ref{scatter}, left panel, we plot the solution of the mean intensity for $\omega=0.4$, $g=0.7$ normalized to the incident radiation field  for a homogeneous spherical cloud with an optical depth to the center $\tau_c=20$ (black curve). The dashed blue curve shows the solution for the same albedo, but for isotropic scattering. This represents the solution of the MCDRT code. The dotted red curve gives the result for pure forward-scattering, i.e., the case $\exp(-(1-\omega)\tau)$. This is equivalent to the previous approximation in \ktau.
The right panel shows the corresponding error when assuming isotropic or pure forward-scattering. It is apparent that the assumption of isotropic scattering is an improvement over our previous approach up to optical depths of about 16, only for very high $\tau$ the forward-scattering is the solution closer to the $g=0.7$ case. However, the FUV influence at $\tau>10$ is negligible. Presently, we therefore sacrifice the dark cloud accuracy in favor of a more accurate description of the cloud surface, but in a future work we plan to update the MCDRT code to properly account for non-isotropic scattering.

\section{Rescaling of photo-rates to account for different dust properties\label{appendix:rescaling}}
\begin{figure*}[htb]
 \centering
 \includegraphics[width=9cm]{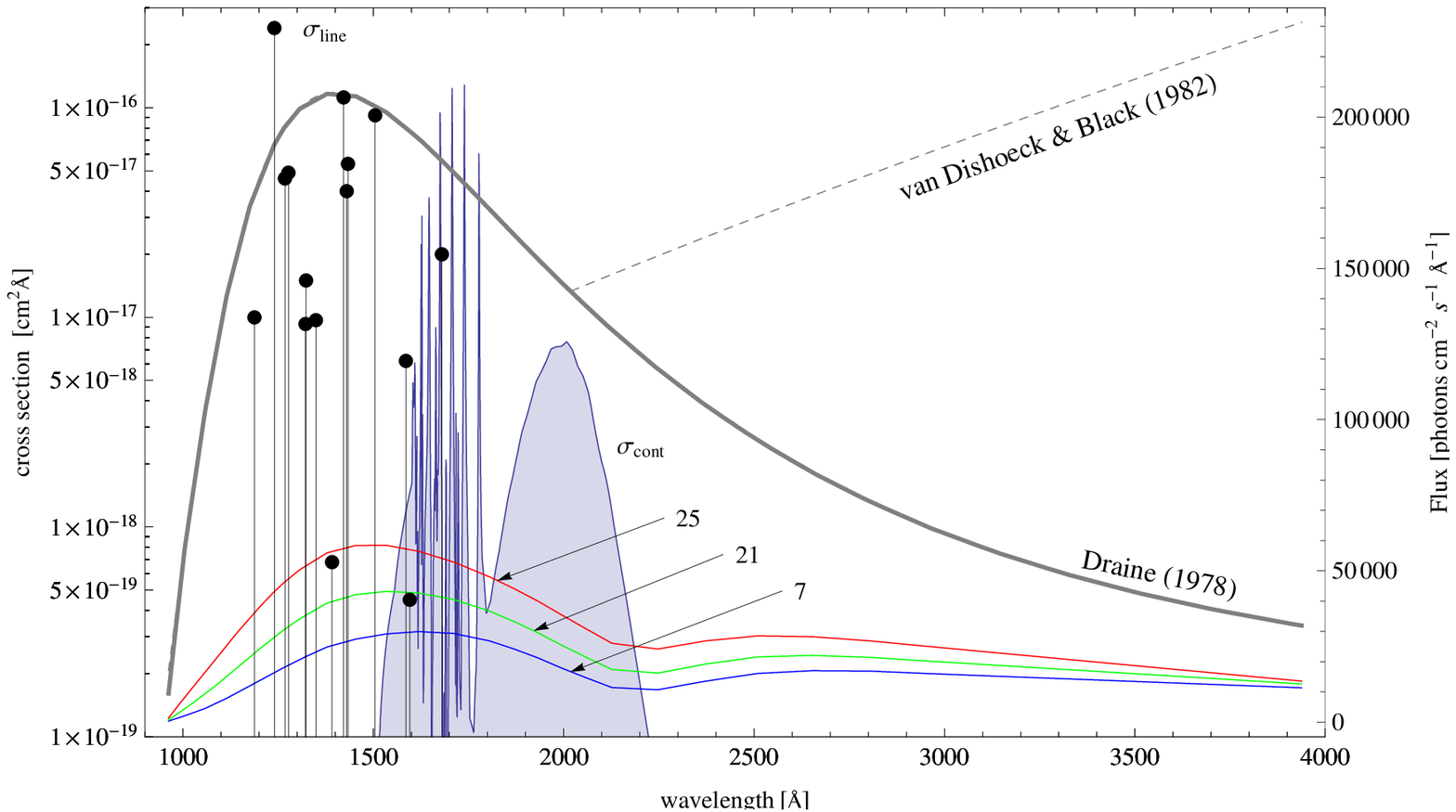}
\hfill
 \includegraphics[width=9cm]{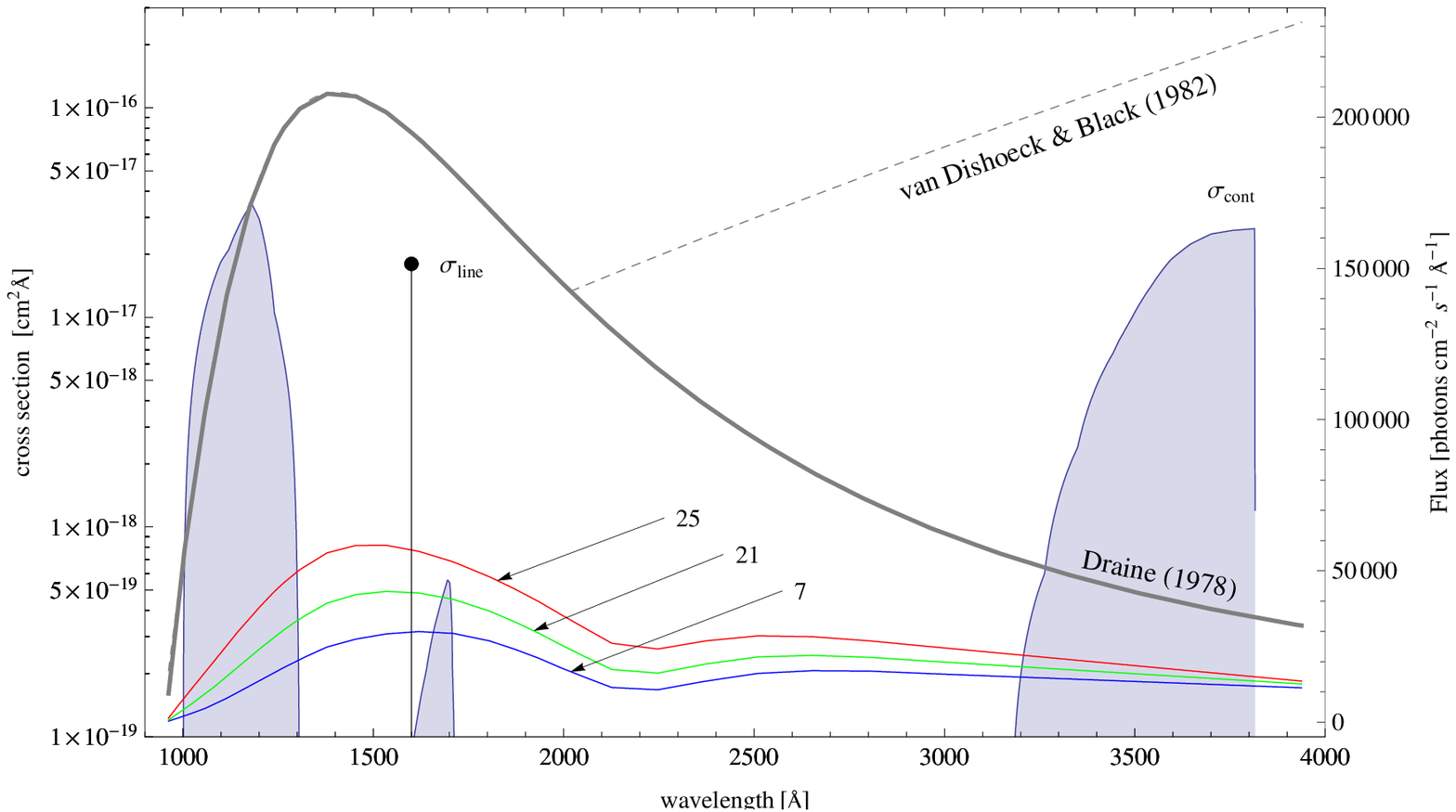}
\caption{Wavelength dependence of the assumed UV illumination and the photo-dissociation cross sections. The thick black line shows the unshielded empty-space Draine FUV field, the dashed line shows an extension of the Draine field longward of 2000 $\AA$ suggested by \citet{vDB82}. The blue, green and red lines show the FUV radiation field inside the reference cloud at 95\%~$R_\mathrm{tot}$ for the WD01-7, WD01-21, and WD01-25 dust models, respectively. {\bf Left panel:}  The blue filled curve shows the continuum absorption cross section $\sigma_\mathrm{cont}$ of \ce{CH2}  that leads to dissociation, the black dots with drop lines show the corresponding line absorption cross section $\sigma_\mathrm{line}$. {\bf Right panel:} Same as left panel but for the \ce{SiH+} photo-dissociation cross sections. } \label{crosssections}
 \end{figure*}
In the following we discuss the uncertainties of the proposed rescaling $\gamma_j\longrightarrow \gamma_j^D$ for different dust models $D$.
 
One important factor is the spectral shape of the assumed FUV radiation field. In Fig.~\ref{crosssections} we plot the photo-dissociation cross sections $\sigma$ of two different molecules (\ce{CH2} in the left panel, \ce{SiH+} in the right panel) together with the FUV radiation field (thick black line: unshielded Draine field, colored lines: FUV field inside the clump at 95\% of the total radius for the dust models WD01-7, 21, and 25). The total photo-dissociation cross section is calculated using Eq.~\ref{photo1}. The different wavelength behavior of $\sigma$ in the two panels demonstrates why the photo-dissociation rates of different species are attenuated differently when going deeper into the cloud. The right panel also emphasizes that the choice of the spectral shape of the illuminating FUV radiation field may influence the photo-dissociation rate of some species significantly; the choice between the two  FUV fields results in an unshielded photo-dissociation rate difference of a factor two. However, differences in the attenuation behavior between different dust models will still be dominated by wavelengths shorter than 2000~\AA\ because $A(\lambda)/A(V)$ varies only weakly for $\lambda>3000\,\AA$.  
\begin{figure}[htb]
\resizebox{\hsize}{!}{\includegraphics{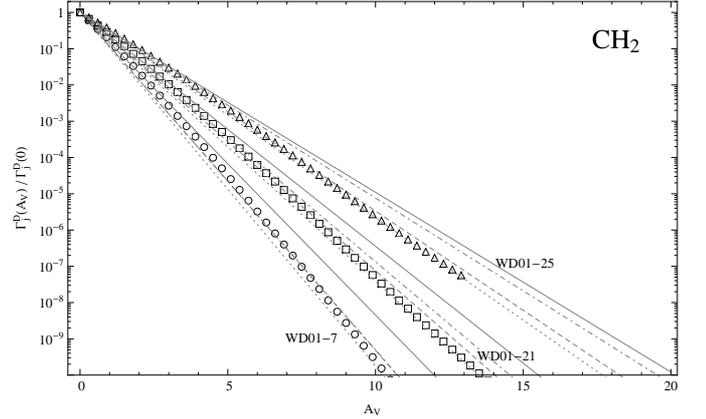}}
\caption{Comparison of the explicitly integrated photo-dissociation rates of \ce{CH2} for the WD01-7, WD01-21, and WD01-25 dust models (circles, squares, triangles, respectively) and the corresponding fits to the rates according to Eq.~\ref{photo2}. The different line styles denote fits up to different maximum visual extinctions (solid: $A_\mathrm{V,max}=3$, dashed: $A_\mathrm{V,max}=10$, dotted: $A_\mathrm{V,max}=15$, and dot-dashed: $A_\mathrm{V,max}=20)$.}
\label{ch2-cross-fit}
\end{figure}

Another uncertainty results from the choice of $A_\mathrm{V,max}$ up to which the $\gamma_{j,fit}$ from Eq.~\ref{photo2} is fitted to the values $\Gamma_j(A_\mathrm{V})/\Gamma_j(0)$. In Fig.~\ref{ch2-cross-fit} we show the $\Gamma_ \ce{CH2}(A_\mathrm{V})/\Gamma_ \ce{CH2}(0)$ for three dust models: WD01-7, WD01-21, and WD01-25. The lines shows the corresponding $\exp(-\gamma_{j,fit}^D A_\mathrm{V})$  fits for different $A_\mathrm{V,max}$. The range of $\gamma_{j,fit}^D$  is 1.92-2.24, 1.48-1.71, and 1.14-1.3 for the three dust models, equivalent to a 10-15\% uncertainty.
We calculated fits to $\gamma_{j,fit}^D$ for all available species and for all dust models to $A_\mathrm{V,max}=10$.

\begin{figure*}[htb]
\centering
\includegraphics[width=17cm]{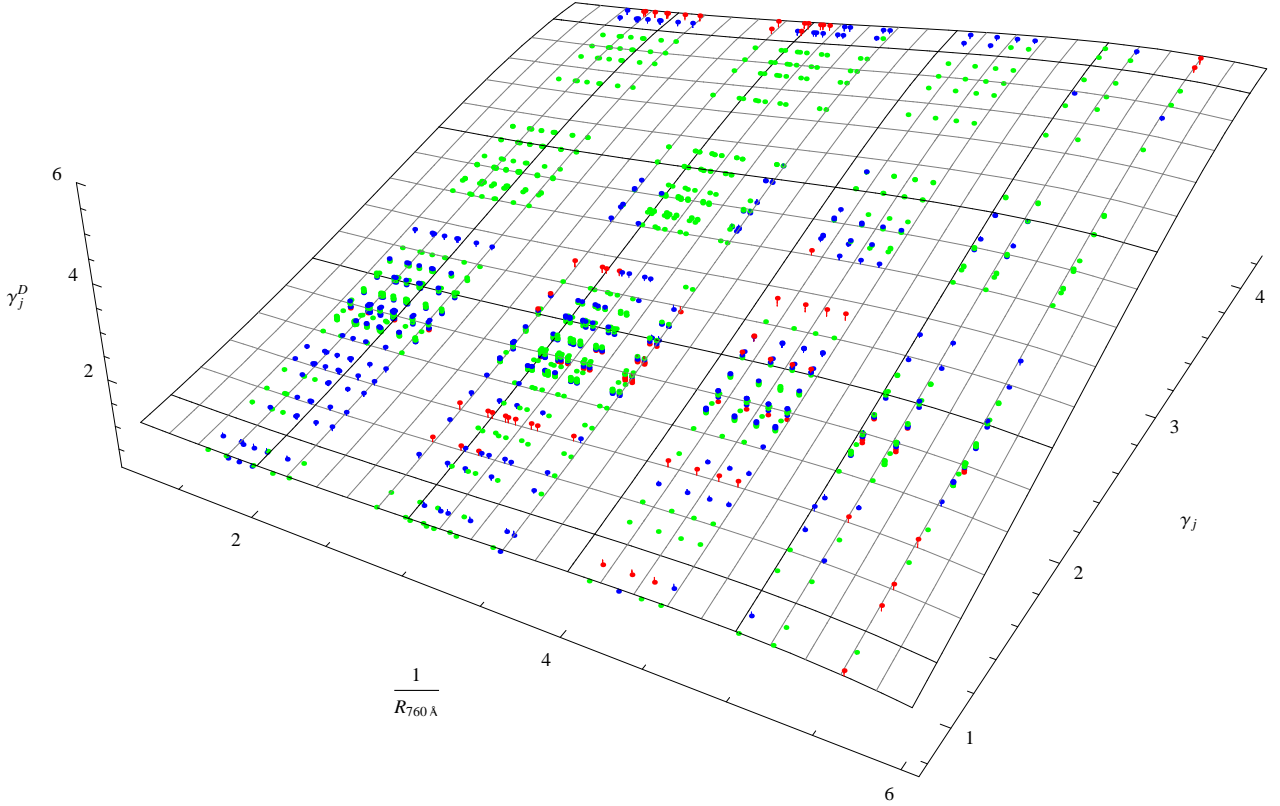}
\caption{Visualization of  Eq.~\ref{gammascaling} (black grid) and the explicitly calculated fits $\gamma_{j,fit}^D$ (colored dots) for the 25 WD01 dust models for all species with available cross sections. The data points are colored according to their deviation from Eq.~\ref{gammascaling}. Green and blue denotes deviations of less than 0.05 and 0.1. Red points signal deviations exceeding 0.1. All deviations are smaller than 0.22.}
\label{rescalingplane}
\end{figure*}

The third source of uncertainties is the functional dependence used in  the fit through Eq.~\ref{gammascaling}.
We performed least-squares fits to polynomials $\gamma_{j}^D=p(\gamma_j, 1/R_{760 \AA}^D)$ up to order 4 to derive a suitable scaling relation between the input parameters $\gamma_j$, describing the wavelength-dependent attenuation of the photo-dissociation rate of species $j$ with $A_\mathrm{V}$,  $1/R_{760 \AA}^D$, describing the  wavelength-dependent attenuation of the FUV radiation field depending on the dust model $D$, and the output parameter $\gamma_{j}^D$. Figure~\ref{rescalingplane} visualizes Eq.~\ref{gammascaling} (black grid) and the explicitly calculated fits $\gamma_{j,fit}^D$ (colored dots) for the 25 WD01 dust models for all species with available cross sections. The data points are colored according to their deviation from Eq.~\ref{gammascaling}.  

Seven species, \ce{CH+}, \ce{SH+}, \ce{OH+}, \ce{HCO+}, \ce{CO}, \ce{O2+}, and \ce{SiO}, deviate significantly from our Eq.~\ref{gammascaling} with residual >0.4. By shifting their $\gamma_j$ to a corrected $\hat{\gamma}_j$ we are able to compensate  for these deviations. Figure~\ref{outlier-rescaling} shows the fitted $\gamma_{j,fit}^D$ of the three strongest outliers and the corresponding $\gamma_j^D$ from Eq.~\ref{gammascaling} using the original $\gamma_j$ (dashed lines) and the corrected $\hat{\gamma}_j$ (solid lines).

The deviations $|\gamma_{j,fit}-\gamma_j^D|$ of the seven outlying species are large compared to the rest of the 63 species that we calculated (compare Fig.~\ref{rescalingplane}). In the following we will discuss the strongest outliers  individually\footnote{We ignore deviations for \ce{CO} because photo-dissociation of \ce{CO} is treated separately in any PDR code \citep{vDB88}}.
\subsubsection*{\ce{HCO+}} 
\ce{HCO+} showed the strongest maximum deviation of 1.12 from the plane defined by Eq.~\ref{gammascaling}. UDfA gives $\gamma=2.0$ referring to \cite{vdishoeck2006} who give a value of $\gamma=3.32$ in their Tab.~2. Unfortunately, all photo-dissociation rate coefficients from UDfA that refer to \cite{vdishoeck2006} are inconsistent with their published numbers and should be replaced with more recent numbers.
Their paper was still in preparation when \citet{udfa06} was published, so the mismatch is possibly due to pre-publication updates of the final values. 
\subsubsection*{\ce{OH+}} 
\ce{OH+} showed the second-strongest deviations of up to 0.8 from Eq.~\ref{gammascaling}. It possesses a strong absorption continuum only for wavelengths shorter than 1100~$\AA$ so that extensions to the Draine FUV field at longer wavelengths should not affect the final photo-dissociation rate. However, computing the unshielded photo-dissociation rate coefficient $\alpha$ for a Draine field yields $1.27\times 10^{-11}$~s$^{-1}$ , comparable to $1.1\times 10^{-11}$~s$^{-1}$ from \citet{vdishoeck2006}, while UDfA lists $1\times 10^{-12}$~s$^{-1}$. The reason for the mismatch is unclear, all sources refer to the same photo-dissociation cross section by \citet{saxon1986}. We also find inconsistent values of $\gamma$ published, between 1.8 and 3.5. We find a best fit to Eq.~\ref{gammascaling} with $\hat{\gamma}_j^D=2.76$.
\subsubsection*{\ce{CH+}} 
For \ce{CH+} deviations from Eq.~\ref{gammascaling} of up to 0.6 were found. \citet{udfa06} give $\alpha=2.5\times 10^{-10}$~s$^{-1}$ and $\gamma=2.5$ referring to \citet{roberge1991}, \citet{vdishoeck2006} give $\alpha=3.3\times 10^{-10}$~s$^{-1}$ and $\gamma=2.94$ using cross sections from \citet{kirby1980}. We find $\gamma_{j,fit}^D=1.5 - 2.4$ and $\hat{\gamma}_j^D=2.11$. At $A_\mathrm{V}$ these lower values of $\gamma$ lead to photo-dissociation rates  3-6 orders of magnitude stronger compared to rates calculated with the higher $\gamma$ from above. The reason for the $\gamma$ offsets is unclear.
\subsubsection*{\ce{SH+}} 
{\ce{SH+} has comparable values of $\alpha$, and $\gamma$ in UDfA and in \citet{vdishoeck2006}. Small differences in $\alpha$ can be attributed to a strong absorption line at 3100~$\AA$, which leads to a stronger unshielded photo-dissociation rate when applying a 10000~K black-body radiation field as used by \citet{udfa06}. We find  $\gamma_{j,fit}^D=1.2 - 1.8$ and $\hat{\gamma}_j^D=1.39$. Differences of $\gamma$ from the literature and our calculations most likely result from different assumed dust properties and fits to lower maximum$A_\mathrm{V,max}$.
\subsubsection*{\ce{SiO} and \ce{O2+}}
Both molecules show maximum deviations of 0.4 from  Eq.~\ref{gammascaling}. The unshielded \ce{SiO} rate coefficients from UDfA and \citet{vdishoeck2006} differ by a factor 16 (see comment on \ce{HCO+}) while their $\gamma_j$ are the same. 
\subsubsection*{}  
In Fig.~\ref{histogram} we show a histogram of the residuals $\gamma_{j,fit}^D-\gamma_j^D$ after removing the outlying species. The standard deviation is $\sigma=0.05$, the maximum absolute residual is 0.22.  
Photo-dissociation cross sections are available for a number of species that are not yet included in UDfA \citep{vhemert2008}. In Tab.~\ref{table:missing} we 
give $\alpha$ and $\gamma$ from http://home.strw.leidenuniv.nl/$\sim$ewine/photo as well as our $\hat{\gamma}_j$.
\begin{figure}[htb]
\resizebox{\hsize}{!}{\includegraphics{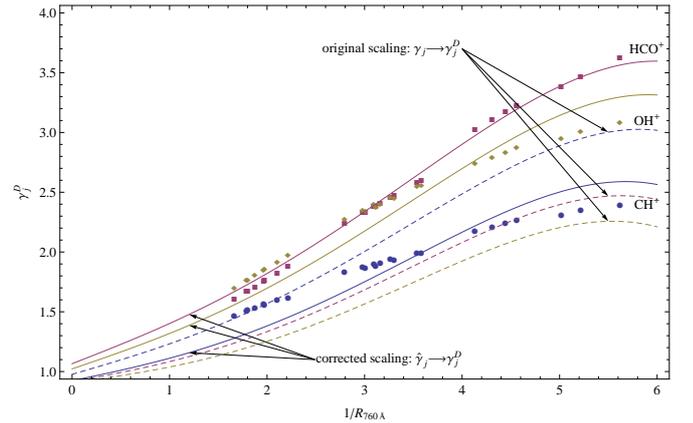}}
\caption{Comparison of the $\gamma_{j,fit}^D$ for the three strongest outliers \ce{CH+}, \ce{OH+}, and \ce{HCO+} and the corresponding scaling behavior of Eq.~\ref{gammascaling} using the original $\gamma_j$ (dashed line) and the corrected $\hat{\gamma}_j$,}
\label{outlier-rescaling}
\end{figure}

\begin{figure}[htb]
\resizebox{\hsize}{!}{\includegraphics{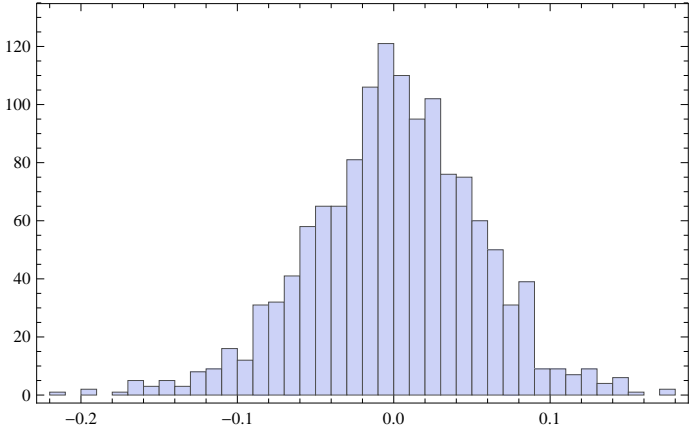}}
\caption{Histogram of the residuals $\gamma_{j,fit}^D-\gamma_j^D$ after removing outliers. 
}
\label{histogram}
\end{figure}

 \begin{table}[htb]
 \begin{center}
 \caption{Rate coefficients and corrected $\hat{\gamma}_j$ for species not  included in UDfA. PI and PD denotes photo-ionization and photo-dissociation.}
 \label{table:missing}
 \begin{tabular}{l l l l l}
 \hline\hline
 process&species&$\alpha$ (s$^{-1}$)&$\gamma_j$&$\hat{\gamma}_j$\\ \hline
 {PI }&\ce{CH4} & $6.8\times 10^{-12}$ & 3.94 & 4.13 \\
 {PD }&\ce{CS2} & $6.1\times 10^{-9}$ & 2.06 & 1.75 \\
 {PI }&\ce{CS2} & $1.7\times 10^{-9}$ & 3.16 & 2.68 \\
 {PD }&\ce{HO2} & $6.7\times 10^{-10}$ & 2.12 & 1.81 \\
 {PD }&\ce{N2O} & $1.9\times 10^{-9}$ & 2.44 & 2.02 \\
 {PI }&\ce{N2O} & $1.7\times 10^{-10}$ & 3.93 & 4.06 \\
 {PD }&\ce{NH+} & $5.4\times 10^{-11}$ & 1.64 & 1.32 \\
 {PI }&\ce{NO2} & $1.5\times 10^{-10}$ & 3.33 & 2.86 \\
 {PD }&\ce{O3} & $1.9\times 10^{-9}$ & 1.85 & 1.48 \\ \hline
\end{tabular}
\end{center}
\end{table}

\section{\HH\ formation rate\label{appendix:h2formation}}
We followed the model of \HH\ formation on interstellar dust grains via physisorption and chemisorption from Cazaux\&Tielens \citep{cazaux2002,cazaux2004, cazaux2010erratum}.

For a given dust type, the total \HH\ formation rate is given by equation \ref{h2rate} (omitting the subscript index $i$ in the following)
\begin{equation}
R_{\mathrm{H}_2}=\frac{1}{2}n_\mathrm{H} v_\mathrm{H} n_d \sigma_d\epsilon_{\mathrm{H}_2}S_{\mathrm{H}}\, , \nonumber
\end{equation}
where the formation efficiency is given by equation \ref{h2form}:
\begin{align}
\epsilon_{\mathrm{H}_2}& =\left(\mathpzc{A}+1+\mathpzc{B}
  \right)^{-1}\xi \nonumber \\
\epsilon_{\mathrm{H}_2}& =\left(\frac{\mu F}{2 \beta_{\mathrm{H}_2}}+1+\frac{\beta_{\mathrm{H}_P}}{\alpha_{\mathrm{P}_C}}\right)^{-1}\xi\, . \label{h2form2}
\end{align}
We set $(\mu F)/(2 \beta_{\mathrm{H}_2})$ to zero to make sure that newly formed \HH\ molecules are able to leave very cold dust surfaces, equivalent to equation (13) in \citet{cazaux2002}. We can approximate $\xi$ and $\beta_{\mathrm{H}_P}/\alpha_{PC}$ by
\begin{equation}
\label{xi}
{\xi=1+\left[\nu_{\mathrm{H}_C}\exp{\left(\frac{1.5 E_{\mathrm{H}_C}}{T_d}\right)}\left(1+\sqrt{\frac{E_{\mathrm{H}_C}-E_S}{E_{\mathrm{H}_P}-E_S}}\right)^2\right]/(2 F)^{-1}}.
\end{equation}
\begin{align}
\label{highTterm}
&\frac{\beta_{\mathrm{H}_P}}{\alpha_{\mathrm{P}_C}}= 
\left(4 \exp\left(\frac{E_S}{T_d}\right)
   \sqrt{\frac{E_{\text{H}_P}-E_S}{E_
   {\text{H}_C}-E_S}}+
\right. \\ \nonumber
&+\left. 
\frac{8 \sqrt{\pi\, T_d}\,\exp\left({-2 a
   \sqrt{\frac{2 m
   \left(E_{\text{H}_P}-E_S\right)}{\hbar
   ^2}}}\right) \exp\left(\frac{E_{\text{H}_P}}{T_d} \right) 
   \sqrt{E_{\text{H}_C}-E_{\text{H}_P}}}{E_{\text{H}_C}-E_S}\right)^{-1} \, .
\end{align}
Equation~\ref{highTterm} takes into account the corrections from \citet{cazaux2010erratum}.
Table (\ref{dustenergies}) gives the applied model parameters for silicate and graphite surfaces.
\begin{table}
\caption{Model parameter for silicate and graphite surfaces. For more details about the determination and calculation of these parameters, see \citet{cazaux2002}}
\label{dustenergies}
\centering
\begin{tabular}{lll}
\hline \hline
Parameter&silicate&graphite\\ \hline
$E_{\mathrm{H}_2}\, [\mathrm{K}]$\tablefootmark{a}&320&520\\
$\mu$\tablefootmark{b}&0.3&0.4\\
$E_S\, [\mathrm{K}]$\tablefootmark{c}&110&260\\
$E_{\mathrm{H}_P}\,[\mathrm{K}]$\tablefootmark{d}&450&800\\
$E_{\mathrm{H}_C}\,[\mathrm{K}]$\tablefootmark{e}&30000&30000\\
$a\sqrt{\frac{2m(E_{\mathrm{H}_P}-E_S)}{\hbar^2}}$\;\tablefootmark{f} &14.4&14.0\\
$\nu_{\mathrm{H}_2}\,[\mathrm{s}^{-1}]$\tablefootmark{g}&$3\times 10^{12}$&$3\times 10^{12}$\\ 
$\nu_{\mathrm{H}_C}\,[\mathrm{s}^{-1}]$\tablefootmark{h}&$1.3\times 10^{13}$&$1.3\times 10^{13}$\\
$F\, [mLy \mathrm{s}^{-1}]$\tablefootmark{i}&\multicolumn{2}{c}{$10^{-10}$}\\
\hline
\end{tabular}
\tablefoot{
\tablefoottext{a}{Desorption energy of \HH.}
\tablefoottext{b}{Fraction of newly formed \HH\ that stays on the surface.}
\tablefoottext{c}{Energy of the saddle point between a chemisorbed and a \mbox{physisorbed} site.}
\tablefoottext{d}{Desorption energy of physisorbed H.}
\tablefoottext{e}{Desorption energy of chemisorbed H.}
\tablefoottext{f}{Product of the width and height of the energy barrier between a physisorbed and a chemisorbed site. For more details see \citet{cazaux2004}}
\tablefoottext{g}{vibrational frequency of \HH\ in surface sites.}
\tablefoottext{h}{vibrational frequency of H in surface sites.}
\tablefoottext{i}{Flux of H atoms in monolayers per second.}
}
\end{table}

\section{Photoelectric heating rates\label{appendix:PEH}}

\begin{figure}
\resizebox{\hsize}{!}{\includegraphics{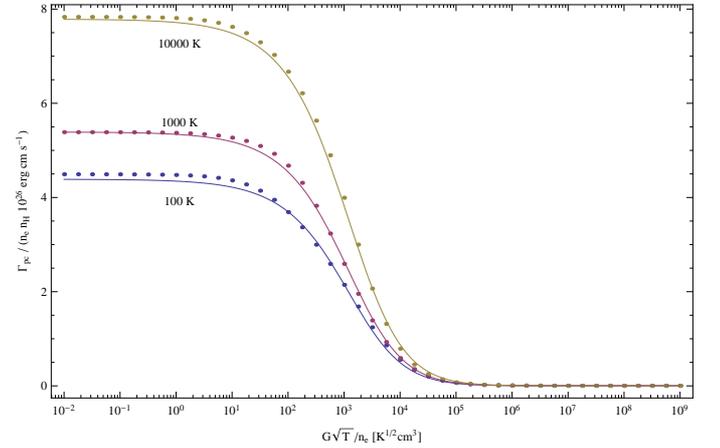}}
\caption{Fit results to the photoelectric heating rate, calculated for an extended range of $G\sqrt{T}/n_e$. The different lines show results for gas temperatures 100~K, 1000~K, and 10000~K. $\Gamma_\mathrm{pc}$ is given in units of $10^{-26}\mathrm{erg}\mathrm{s}^{-1}$. }
\label{PEHappendix}
\end{figure}
\begin{figure}
\resizebox{\hsize}{!}{\includegraphics{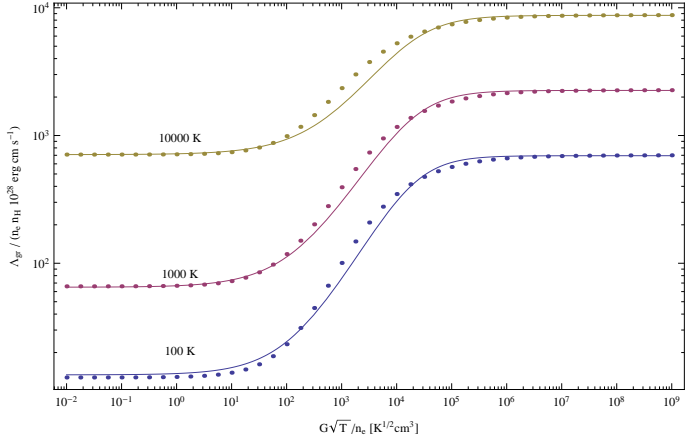}}
\caption{Fit result to the net cooling rate due to collisions with charged particles, calculated for an extended range of $G\sqrt{T}/n_e$. The different lines show results for gas temperatures 100~K, 1000~K, and 10000~K. $\Gamma_\mathrm{pc}$ is given in units of $10^{-26}\mathrm{erg}\mathrm{s}^{-1}$. }
\label{ggappendix}
\end{figure}

\begin{table*}[htb]
\begin{center}
\caption{Photoelectric heating parameters (equation \ref{PEHnew}). See \citet{WD01PEH} for detailed descriptions of the dust size distributions and applied physical parameters.}\label{tab1}
\begin{tiny}
\begin{tabular}{llcccccccccc}
\hline \hline
$R_V$&$b_C$&Distr.&Rad. Field& $C_0$&$C_1$&$C_2$&$C_3$&$C_4$&$C_5$&$C_6$&$C_7$\\\hline
3.1&6.0&A&B0&  8.36780 & 0.20755 & 0.00969 & 0.01354 & 0.41047 & 0.62072 & 0.53099 & 0.00589 \\
4.0&4.0&B&B0& 5.67140 & 0.13563 & 0.00994 & 0.01568 & 0.41122 & 0.60659 & 0.53265 & 0.00382  \\
5.5&3.0&B&B0& 5.67140 & 0.13563 & 0.00994 & 0.01568 & 0.41122 & 0.60659 & 0.53265 & 0.00382 \\\hline
\end{tabular}
\end{tiny}
\end{center}
\end{table*}

\begin{table*}[htb]
\begin{center}
\caption{Collisional cooling parameters (equation \ref{lamdanew}). See \citet{WD01PEH} for detailed descriptions of the dust size distributions and applied physical parameters.}\label{tab2}
\begin{tiny}
\begin{tabular}{llllllllllllllll}
\hline \hline
$R_V$&$b_C$&Distr.&Rad. Field& $D_0$&$D_1$&$D_2$&$D_3$&$D_4$&$D_5$&$D_6$&$D_7$&$D_8$&$D_9$&$D_{10}$&$D_{11}$\\\hline
3.1&6.0&A&B0& 0.493 & 0.556 & 0.031 & 0.842 & 287 & 0.360 & 3.640 & -0.098 & 0.642 &$ 3.25\times 10^{-6}$ & 2.080 & -1.500 \\
4.0&4.0&B&B0& 0.493 & 0.362 & 0.038 & 0.803 & 287 & 0.360 & 2.460 & -0.035 & 0. &$ 2.35\times 10^{-5}$ & 1.870 & -1 \\
5.5&3.0&B&B0& 0.484 & 0.720 & 0.118 & 0.597 & 100 & 0.710 & 1.240 & -0.030 & 0.603 &$ 1.11\times 10^{-6}$ & 2.150 & -2.720\\ \hline
\end{tabular}
\end{tiny}
\end{center}
\end{table*}

\citet{WD01PEH} showed that the photoelectric heating rate for the WD01 grain size distributions is fairly well reproduced by the parametrization
\begin{eqnarray}\label{PEHold}
\Gamma_\mathrm{pc}=G n_\mathrm{H}\frac{C_0+C_1 T^{C_4}}{1+C_2\psi^{C_5}[1+C_3\psi^{C_6}]} 10^{-26}\,\mathrm{erg}\,\mathrm{s}^{-1}\,\mathrm{cm}^{-3},
\end{eqnarray}
with $\psi=(G\sqrt{T})/n_e$, where $T$ is the gas temperature in Kelvins and $\psi$ is in units of K$^{1/2}$cm$^3$, and $G=1.71 \chi$.  Numerical values for the parameters $C_i$ in Eq. (\ref{PEHold}) are given in Table 2 in \citet{WD01PEH}  for each dust distribution. The rate of cooling due to charged particle collisions can be approximated by
\begin{eqnarray}
\label{lambdaold}
\Lambda_\mathrm{gr}&=&n_e n_\mathrm{H} T^{D_0+D_1/x}\exp{D_2+D_3 x-D_4 x^2}\\  \nonumber
& & \times  10^{-28} \,\mathrm{erg}\,\mathrm{s}^{-1}\,\mathrm{cm}^{-3},
\end{eqnarray}
where $x=\ln{(G\sqrt{T}/n_e)}$. Values for the $D_i$ for all WD01 grain distributions are given in Table 3 in \citet{WD01PEH}. 

Equations (\ref{PEHold}) and (\ref{lambdaold}) are fairly accurate when $10^3\mathrm{K}\le T\le 10^4 K$ and $10^2\mathrm{K}^{1/2}\mathrm{cm}^3\le \psi\le 10^6 \mathrm{K}^{1/2}\mathrm{cm}^3$.
However, the parameter range in PDR models allows  $\psi$ to  reach values as high as $10^9 \mathrm{K}^{1/2}\mathrm{cm}^3$ and as low as $10^{-2} \mathrm{K}^{1/2}\mathrm{cm}^3$. \citet{weingartner09} provided updated calculations of $\Gamma_\mathrm{pc}$ and $\Lambda_\mathrm{gr}$ for the dust distributions WD01-7, WD01-21, and WD01-25 up to  $\psi=10^9 \mathrm{K}^{1/2}\mathrm{cm}^3$.

We performed numerical fits to the updated photoelectric heating rates and to the cooling rate due to collisions with charged particles according to the parametrizations \ref{PEHnew} and \ref{lamdanew}. The heating and cooling rates have been calculated by \citet{weingartner09} with an extended parameter range of $10^{-2}\mathrm{K}^{1/2}\mathrm{cm}^3\le \psi\le 10^9 \mathrm{K}^{1/2}\mathrm{cm}^3$. The details of the computation are described in \citet{WD01PEH}. To achieve a good fit over the full extended parameter range it was necessary to modify the algebraic form of the parametrizations relative to its original form in \citet{WD01PEH}. The final form is given in equations  \ref{PEHnew} and  \ref{lamdanew}. In Figs. \ref{PEHappendix} and \ref{ggappendix} we show the fits to calculations of $\Gamma_\mathrm{pc}$ and $\Lambda_\mathrm{gr}$ for gas temperatures of 100~K, 1000~K and 10000~K to demonstrate the quality of the parametrization. The $C_i$ and $D_i$ parameters are given in Table \ref{tab1} and \ref{tab2} and will provide a fairly accurate reproduction of the total photoelectric heating rate $\Gamma_\mathrm{tot}$ between $10~K\le T \le 10000 K$ and  $10^{-2}\mathrm{K}^{1/2}\mathrm{cm}^3\le \psi\le 10^9 \mathrm{K}^{1/2}\mathrm{cm}^3$.

\end{document}